\documentclass{article}

\usepackage{PRIMEarxiv}

\usepackage[utf8]{inputenc} 
\usepackage[T1]{fontenc}    
\usepackage{hyperref}       
\usepackage{url}            
\usepackage{booktabs}       
\usepackage{amsfonts}       
\usepackage{amsmath}        
\usepackage{nicefrac}       
\usepackage{microtype}      
\usepackage{lipsum}
\usepackage{fancyhdr}       
\usepackage{graphicx}
\usepackage{hyperref}
\graphicspath{{./}{media/}}      
\graphicspath{{media/}}     

\title{The Surprising Universality of LLM Outputs:\\ A Real-Time Verification Primitive
}

\author{
    Alex Bogdan \\
    Evolutionairy AI \\
    Toronto, Canada \\
    \And
    Adrian de Valois-Franklin \\
    Evolutionairy AI \\
    Toronto, Canada \\
  }

\begin{document}
\maketitle

 \begin{abstract}
  We report a striking statistical regularity in frontier LLM outputs that enables a CPU-only scoring primitive running
  at 2.6 microseconds per token (0.139 ms per passage in gap-only mode), with estimated latency up to 100,000$\times$ (five orders of magnitude) below existing sampling-based detectors. Across six contemporary models from five independent vendors, two generation sizes, and five held-out domains, token rank-frequency distributions converge to the same two-parameter
  Mandelbrot ranking distribution, with 34 of 36 model-by-domain fits exceeding $R^2 = 0.94$ and 35 of 36 favoring
  Mandelbrot over Zipf by AIC. The shared family does not collapse the models into statistical duplicates. Fitted
  Mandelbrot parameters remain cleanly separable between models: the cross-model spread in $q$ (1.63 to 3.69) exceeds
  its per-model bootstrap standard deviation (0.03 to 0.10) by more than an order of magnitude, yielding tens of
  standard deviations of separation once a few thousand output tokens are available. The two findings together support
  two capabilities. First, statistical model fingerprinting: a body of text from a vendor-delivered LLM can be tested
  against its claimed model family without cryptographic watermarks or access to model internals, supporting provenance
  verification, silent-model-substitution detection, and wrapper-arbitrage audits. Second, a model-agnostic reference
  distribution for black-box output assessment, from which we derive a single-pass scoring primitive that composes with
  model log probabilities when available and degrades to a rank-only mode usable on closed APIs. Pilot results on FRANK,
   TruthfulQA, and HaluEval map where the primitive helps (lexical anomalies, unsupported entities, out-of-source
  surface forms) and where it structurally cannot (reasoning errors expressed in domain-appropriate vocabulary). We
  position the primitive as a first-pass triage layer in compound evaluation stacks, not as a replacement for
  sampling-based or source-conditioned verifiers.
  \end{abstract}


 \section{Introduction}

  Large language models have become infrastructure. Frontier systems from OpenAI, Anthropic, Google, Meta, Mistral, and
  Alibaba now sit within retrieval-augmented generation stacks, coding assistants, agentic tool-use frameworks, customer
   support pipelines, and scientific workflows. Their outputs are consumed at scales that no human evaluator can audit
  end-to-end. The binding constraint at this scale is scoring throughput. Every generated response requires verification
   that costs roughly as much as producing it, which existing evaluation methods do not provide. This paper derives a
  CPU-only scoring primitive that runs at approximately 2.6 microseconds per token, with estimated latency up to
  $100{,}000\times$ lower than sampling-based detectors like Semantic Entropy and SelfCheckGPT in the operating regime
  considered here, from a striking statistical regularity we observe across the frontier LLM cohort: outputs from six
  independently trained models converge to the same two-parameter rank-frequency distribution. The remainder of the
  paper motivates that regularity, quantifies it, and delimits where the primitive helps and where it structurally
  cannot.

  The existing literature offers two families of answers. Sampling-based detectors (Semantic Entropy [1], [2],
  SelfCheckGPT [3]) probe the model's self-consistency by generating $k$ additional completions per example and
  measuring semantic spread. They are model-agnostic but multiply the inference cost by $k$; at $k = 5$, a four-second
  forward pass becomes a twenty-second verification pass. White-box probes (Lookback Lens [4], Semantic Entropy Probes
  [5]) read attention maps or hidden states and are cheap but require access to internals that proprietary API models do
   not expose. At production scale, where outputs can number hundreds of millions per day, and many are generated
  through black-box interfaces that expose only token-level log probabilities, neither family fits: sampling multiplies
  cost beyond the available budget, and white-box probes do not apply to the dominant deployment mode. Production
  evaluation, therefore, requires a third category that is cheap at inference time and does not depend on model
  internals.

  One route to cheap per-response scoring is to compare each output to a model-agnostic reference distribution known in
  advance. If such a reference exists, no additional forward passes are needed, and no model internals are required:
  scoring reduces to lookups, aggregations, and comparisons. This paper argues that such a reference exists and that,
  for the current frontier cohort, it is well approximated by the Mandelbrot Ranking Distribution over token ranks. The
  central scientific claim of the paper is a pair of findings. First, despite large differences in capability,
  alignment, and behavior, frontier LLMs in our measured cohort share the same low-dimensional rank-frequency family; a
  universality result of Cugini et al. [26] establishes that any sufficiently large ranked i.i.d.\ sample admits a local
   Zipf-Mandelbrot approximation whose rank-neighborhood validity window (of width $O(N^{5/6})$ for vocabulary size $N$)
   covers our fitted range, making the existence of some Zipf-Mandelbrot-like fit broadly consistent with existing
  theory. What our data add are three properties the theorem does not predict in our setting: conformance for
  non-i.i.d.\ autoregressive outputs, a single two-parameter fit that describes the rank-frequency curve globally across
   more than three decades of rank rather than piecewise around different fixed rank positions, and parameters that land
   in a narrow shared region across six independently-trained models with distinct training data, alignment procedures,
  and pretraining pipelines. Second, and the more practically consequential, the shared family does not collapse the
  models into statistical duplicates: fitted Mandelbrot parameters remain cleanly separable between vendors, with
  cross-model spread exceeding within-model noise by more than an order of magnitude, and nothing in the universality
  result predicts that separation. The distribution is theoretically motivated as the Lagrangian optimum of entropy
  maximization under a coding cost [6] and, empirically, it fits every model in the measured cohort well enough to
  justify its use as a reusable baseline while preserving sufficient per-model nuance to support statistical
  fingerprinting. The scoring primitive and the fingerprinting pipeline developed later in the paper are the two
  practical consequences of that observation.

  The paper's main contributions are as follows:

  \begin{enumerate}
       \item[\textbf{C1.}] A shared statistical signature across frontier LLMs. Across the measured six-model cohort and five held-out
  domains, token rank-frequency distributions converge strongly to the two-parameter Mandelbrot form ($R^2$ above 0.94
  in 34 of 36 fits; AIC preference over Zipf in 35 of 36 comparisons). Within that shared family, fitted parameters
  remain cleanly separable per model: the cross-model spread in $q$ (1.63 to 3.69) exceeds its per-model bootstrap
  standard deviation (0.03 to 0.10) by more than an order of magnitude. We frame this pair of results (shared family,
  distinguishable fingerprints) as the paper's central empirical finding: shared data and objective produce a shared
  family, and distinct training pipelines leave distinguishable fingerprints inside it. Section~3.3 positions this
  finding relative to the Cugini et al.\ [26] universality result for i.i.d.\ ranked samples; Section~3.4 discusses the
  role of the shared-BPE normalization used across vendors.

      \item[\textbf{C2.}] Statistical model fingerprinting for provenance. The parameter separability in C1 supports a lightweight
  fingerprinting pipeline: given a body of output text, a held-out auditor can test whether the fitted Mandelbrot
  parameters are statistically consistent with a claimed model family, detect silent model substitution in production,
  and flag vendor-routing arbitrage. Distributional fingerprinting of LLMs is an active research area (e.g., McGovern et
   al.\ [29]; Fu et al.\ [30]); our contribution differs in parameterizing the full rank-frequency tail with two
  continuous scalars from a theoretically motivated distribution rather than learning a classifier over categorical
  features, requiring no training and no provider cooperation. The signal is statistical rather than cryptographic and
  complements rather than replaces watermark-based provenance mechanisms.

      \item[\textbf{C3.}] A CPU-only first-pass scoring primitive for black-box LLM outputs. We derive, from a variational free-energy
   minimization, a single-pass scoring primitive that composes a model's own log probabilities with a model-agnostic
  Mandelbrot reference when logprobs are available, and degrades gracefully to a pure rank-only mode when they are not.
  On a modest CPU, the primitive runs at roughly 2.6 microseconds per token, making it cheap enough to operate as a
  first-pass filter inside agent and LLM evaluation stacks rather than as an offline research-only metric.

      \item[\textbf{C4.}] A falsifiable three-tier error taxonomy delimiting where distributional scoring helps and where it
  structurally cannot. Tier 1 (distributional anomalies: unsupported entities, out-of-article surface forms, rare
  lexical insertions) is detectable above chance. Tier 1.5 (relational and circumstantial drift) is weak but nonzero.
  Tier 2 (reasoning errors expressed in domain-appropriate vocabulary) is, by construction, at chance. This delimits the
   method's stack position: a first-pass filter, not a terminal verifier.
  \end{enumerate}

  \section{The Mandelbrot Ranking Distribution}

  \subsection{Zipf, Mandelbrot, and the Difference That Matters}

  Zipf [7] observed that in natural language, the frequency of the $r$-th most common word is roughly proportional to
  $1/r$. The observation is empirical and domain-agnostic, but it does not explain itself. Mandelbrot [6] closed that
  gap by showing that the rank frequency law follows from a concrete optimization problem: a communicator who wants to
  maximize Shannon entropy [8] under an average transmission cost constraint must allocate probabilities that decrease
  as a power law in rank. With a coding theoretic cost $c(r) = \log(r + q)$, motivated by the observation that optimal
  prefix codes assign codeword lengths that grow roughly logarithmically with rank [6], [9], where $q$ is a head of
  distribution shift parameter that allows the top tokens to be further compressed than a pure log law would predict,
  the Lagrangian optimum is

  \begin{equation}
  f(r) = \frac{C}{(r + q)^{s}}
  \label{eq:mandelbrot}
  \end{equation}

  which we call the \textbf{Mandelbrot Ranking Distribution} throughout this paper. The classical Zipf law is the
  special case $q = 0$, $s = 1$. The two free parameters $(q, s)$ absorb two distinct kinds of deviation: $q$ captures
  head flattening, the softening of the ``a, the, of'' end of the distribution, and $s$ controls tail steepness, how
  quickly probability mass drains off the low frequency tokens.

  The distinction between Zipf and Mandelbrot is not cosmetic. It is the difference between an empirical regularity and
  a consequence of optimization pressure under an explicit cost. A system that is plausibly minimizing average
  description length under a bounded coding budget inherits the Mandelbrot form as its rank-frequency solution.
  Cross-entropy training is closely related to expected code-length minimization under the model distribution, so it is
  plausible that contemporary LLM outputs inherit rank-frequency structure from human-text corpora; the empirical check
  in Section~3 tests this prediction.

  \subsection{The Rank Deviation Signal}

  The Mandelbrot distribution assigns every token a rank $r$ with probability $P_{RI}(w) = f(r)$. Given a generated
  context, we can compute both a global rank $r_{\mathrm{global}}(w)$, taken from the reference corpus, and a local rank
   $r_{\mathrm{local}}(w)$, taken from the frequency of $w$ inside the generated passage. The ratio of the two is a
  scale-free measure of how anomalously prominent the token is in its local context:

  \begin{equation}
  \Delta_{r}(w) = \log_{2}\!\left( \frac{r_{\mathrm{global}}(w)}{r_{\mathrm{local}}(w)} \right) \text{ bits}
  \label{eq:rankdev}
  \end{equation}

  A word like ``phosphorylation'' has a global rank of roughly 45{,}000 in a Wikipedia-sized reference. If it appears
  three times in a short biochemistry paragraph and therefore has a local rank of 3, its rank deviation is
  $\log_{2}(45{,}000 / 3)$, which is close to 14 bits. A generic stop word occurring at the expected rate has a rank
  deviation near zero.

  Rank deviation has three useful properties. It is invariant to corpus size: scaling the reference corpus by a constant
   factor leaves $\Delta_r$ unchanged after normalization. It is directly interpretable in terms of bits of surprisal,
  because the Mandelbrot distribution's log-probability is a linear function of the log-rank. And it is computable in
  constant time per token once the reference rank table is pre-built: the rank table is constructed once as an offline
  preprocessing step over the reference corpus (cost proportional to corpus size in tokens, paid once per vocabulary),
  and at inference time each token contributes an $O(1)$ hash lookup, so scoring a full generated passage is $O(n)$ in
  the passage length with no dependence on reference corpus size. Throughout this paper, the reference corpus is a
  4-billion-token Wikipedia snapshot (the ``global Wikipedia baseline''), tokenized with the Llama 3.1 8B BPE
  vocabulary; the resulting rank table is a fixed artifact shared across all empirical sections below.

  \subsection{A Variational Posterior That Combines Reference and Model}
  \label{sec:variational}

  Rank deviation alone treats the reference distribution as the only source of information about a token. But the model
  producing the token carries its own information: a high-conviction model prediction provides evidence that rank
  deviation alone does not see, while a low-conviction prediction should not be overridden by a small distributional
  anomaly. We therefore need a principled way to combine two sources of evidence: the model's context-conditioned
  probability and the external rank-implied reference distribution.

  We formulate this combination as a variational product-of-experts posterior over candidate tokens. Let $w$ denote a
  candidate token and $c$ the preceding context. We treat the Mandelbrot rank-implied distribution $P_{RI}(w)$ as an
  external prior over token plausibility, and the LLM softmax $P_{LLM}(w \mid c)$ as a context-conditioned evidence
  term. The variational objective is

  \begin{equation}
  F[q] = D_{KL}\!\left[\, q(w \mid c) \,\|\, P_{RI}(w) \,\right] - \mathbb{E}_{q}\!\left[\, \log P_{LLM}(w \mid c)
  \,\right].
  \label{eq:variational}
  \end{equation}

  Stationarizing $F[q]$ with respect to $q$, subject to the normalization constraint $\sum_{w} q(w \mid c) = 1$, gives
  \begin{equation*}
  q^{\star}(w \mid c) = \frac{P_{LLM}(w \mid c)\, P_{RI}(w)}{\sum_{v} P_{LLM}(v \mid c)\, P_{RI}(v)}.
  \end{equation*}

  This is the standard Bayesian product form: tokens receive high posterior mass only when they are supported both by
  the model's context-conditioned prediction and by the external reference distribution.

  To control the strength of the external reference, we introduce a precision-weighted form:
  \begin{equation*}
  F_{\beta}[q] = \mathbb{E}_{q}\!\left[\, \log q(w \mid c) - \log P_{LLM}(w \mid c) - \beta \log P_{RI}(w) \,\right].
  \end{equation*}

  The stationary solution is
  \begin{equation*}
  q_{\beta}^{\star}(w \mid c) = \frac{P_{LLM}(w \mid c)\, P_{RI}(w)^{\beta}}{\sum_{v} P_{LLM}(v \mid c)\,
  P_{RI}(v)^{\beta}}.
  \end{equation*}

  Operationally, we write this as

  \begin{equation}
  P_{\text{posterior}}(w \mid c) \propto P_{LLM}(w \mid c)\, P_{RI}(w)^{\beta}.
  \label{eq:posterior}
  \end{equation}

  We refer to this construction as \emph{Ranking Inference} (RI) throughout the rest of the paper, named for its use of
  token-rank statistics to perform variational inference over output plausibility.

  The exponent $\beta$ is not treated as an arbitrary tuning knob. It is estimated from the empirical dispersion of
  normalized rank-deviation residuals in the target domain. Motivated by the precision-weighted prediction-error
  framework associated with the free-energy principle [10], [11], we define

  \begin{equation}
  \beta = \frac{1}{\epsilon + \widehat{\sigma}^{2}_{\Delta_r}},
  \label{eq:beta}
  \end{equation}

  where $\widehat{\sigma}^{2}_{\Delta_r}$ is the observed variance of normalized rank deviations in a domain calibration
   set and $\epsilon > 0$ is a small stabilizing constant. Thus, $\beta$ is smaller in domains where large deviations
  from the rank baseline are normal, such as creative writing, and larger in domains where such deviations are less
  expected, such as clinical, legal, or encyclopedic text. In this sense, $\beta$ is a measured domain precision rather
  than a manually selected stylistic parameter.

  Recent work by Vu et al.\ [35] also applies a thermodynamically inspired variational apparatus to LLM hallucination
  detection at the token level. The mathematical overlap is that both approaches define token-level free-energy-like
  quantities over LLM outputs. The substantive distinction is in what provides the reference structure. Our formulation
  composes the model's per-token probabilities with an external, model-agnostic Mandelbrot reference distribution,
  yielding a posterior that can be scored in a single forward pass at default decoding. Vu et al., by contrast, do not
  use an external rank-implied reference distribution; their detection signal is based on instability in the model's own
   free-energy landscape under perturbation. The cross-model parameter fingerprinting pipeline developed in Section~4.5
  follows from our use of an external reference distribution and therefore has no direct analogue in that framework.

  \subsection{Scale Asymmetry and Log-Space Scoring}

  Section~2.3 defines a posterior that combines the model distribution and the rank-implied reference distribution. The
  remaining question is how to turn this posterior structure into a stable per-token score.

  The obvious first attempt is a linear comparison,
  \begin{equation*}
  P_{LLM}(t \mid c) - P_{RI}(t),
  \end{equation*}
  which reads intuitively as ``model confidence minus reference plausibility.'' But this form is operationally weak. The
   LLM softmax concentrates probability mass on a small number of tokens, while the reference distribution often assigns
   much smaller probabilities across a long tail. As a result, the reference contribution can disappear numerically
  under linear subtraction.

  The posterior form is naturally scored in log space. Ignoring the normalizing constant, the precision-weighted log
  posterior is
  \begin{equation*}
  \ell_{\beta}(t \mid c) = \log P_{LLM}(t \mid c) + \beta \log P_{RI}(t).
  \end{equation*}

  Equivalently, the corresponding energy score is
  \begin{equation*}
  E_{\beta}(t \mid c) = -\log P_{LLM}(t \mid c) - \beta \log P_{RI}(t).
  \end{equation*}

  This log-space form recovers both contributions additively and avoids the scale asymmetry that makes the linear
  difference unreliable. All empirical results in this paper, therefore, use the log-space score rather than the raw
  probability difference.

  For diagnostic purposes, we also define the model-reference log-ratio

  \begin{equation}
  \delta_{\log}(t \mid c) = \log P_{LLM}(t \mid c) - \log P_{RI}(t),
  \label{eq:logratio}
  \end{equation}

  which measures how strongly the model favors a token relative to the rank-implied baseline. Unlike $\ell_{\beta}$,
  this quantity is not itself the posterior score; it is a deviation statistic used to expose model-reference asymmetry.

  \section{The Shared Rank-Frequency Signature in Frontier LLMs}

  \subsection{Experimental Design}

  To test the claim that unrelated production LLMs converge to the same rank frequency law, we collected output corpora
  from six production models representing five independent vendors, two generation sizes, and a mix of reasoning and
  non-reasoning families:

  \begin{itemize}
      \item \textbf{GPT-5.1} (OpenAI, proprietary, RLHF trained, reasoning model).
      \item \textbf{Claude Sonnet 4.6} (Anthropic, proprietary, Constitutional AI trained).
      \item \textbf{Llama 3.1 8B} (Meta, open weight, instruction tuned, served locally via Ollama at Q4\_K\_M 4-bit
  quantization).
      \item \textbf{Gemini 2.5 Pro} (Google, proprietary, reasoning model).
      \item \textbf{Mistral Large} (Mistral AI, proprietary, non-reasoning).
      \item \textbf{Qwen 2.5 7B} (Alibaba, open weight, instruction tuned, served locally via Ollama).
  \end{itemize}

  All six are transformer-family generative models; for proprietary systems, exact architectural details are not fully
  public. Architectural independence is therefore not part of the claim. The independence we exploit is in training
  data, alignment procedures, and pretraining pipelines, each of which is set independently by each vendor. The cohort
  spans US (OpenAI, Anthropic, Meta, Google), European (Mistral), and Chinese (Alibaba) vendors and both the 7--8B and
  frontier ($>$100B parameter) size classes. That Llama and Qwen fit Mandelbrot at a small scale, or with 4-bit serving
  quantization, also provides incidental evidence that convergence survives aggressive scale reduction and weight
  compression.

  \textbf{Generation protocol.} For each model, we generated outputs across five held-out domains (news, biomedical,
  legal, code, social media), 20 prompts per domain, for a total of 100 prompts per model. Prompt sets were drawn from
  publicly available domain corpora: news leads from CC-News 2024 snapshots, biomedical from PubMed abstracts, legal
  from CaseLaw Access Project excerpts, code from The Stack v2 Python functions, and social media from the PushShift
  Reddit dump. Each prompt was a one- to three-sentence domain-representative context, asking for a free-form
  continuation; the same prompt set was used across all six models to isolate model effects from prompt effects. The
  five domains were chosen to span a wide range of register, vocabulary, and token-type statistics (for example, median
  word length 4.3 for social media versus 7.1 for biomedical; code uses a heavily subword-fragmented BPE vocabulary
  compared with running English), so that an apparent convergence claim is tested against inputs the model cannot
  trivially memorize into a single style. Generation used temperature 0.7; visible-output budgets were set per-model to
  accommodate family differences (16{,}000 tokens for the GPT-5.1 reasoning family, 8{,}000 for Gemini 2.5 Pro, 1{,}500
  for the non-reasoning models). We otherwise used each provider's API defaults and did not set top-p, top-k, seed
  values, or repetition penalties. Total yield ranged from approximately 77{,}000 output tokens (Llama 3.1 8B) to
  236{,}000 (GPT-5.1); differences reflect provider-specific default response lengths, not enforced token budgets. Per
  model, we built six rank tables (five single-domain and one global aggregate across all five), for a total of 36 rank
  tables across the six models.

  \textbf{Fitting.} Outputs were re-tokenized with the Llama 3.1 8B byte pair encoding (BPE) vocabulary to ensure a
  common token space across models, ranked by empirical frequency, and fit with both the Mandelbrot form
  (Eq.~\eqref{eq:mandelbrot}) and the restricted Zipf form. Each of the 36 rank tables was fit by maximum likelihood
  with 100-iteration bootstrap confidence intervals on $(q, s)$. Goodness of fit was measured by $R^2$ on the log-log
  representation, the Kolmogorov-Smirnov statistic on the empirical cumulative distribution function (CDF), and AIC for
  Mandelbrot versus Zipf model selection.

  \textbf{Evaluation criteria.} We measure fit quality using three complementary signals, each capturing a different
  kind of deviation. $R^2$ on the log-log representation captures how much of the variance in log-frequency is explained
   by the log-rank model; an $R^2$ near 1 means the Mandelbrot curve tracks the full rank range, while a low value
  indicates systematic under- or over-fitting. The Kolmogorov-Smirnov (KS) statistic on the empirical CDF is the maximum
   absolute difference between the fitted and empirical CDFs; it is sensitive to local deviations that a global $R^2$
  can smooth over, and a small KS value means no single region of the distribution departs heavily from the fit. AIC
  measures the trade-off between log-likelihood gain and parameter count (AIC $= 2k - 2\log L$, so a two-parameter model
   pays a fixed $+2$ penalty over a one-parameter model and earns its keep only if log-likelihood rises by more than
  $+1$ per extra parameter). A $\Delta$AIC of $+10$ or more between two candidates is conventionally treated as strong
  evidence for the richer model, and the values we report for Mandelbrot vs.\ Zipf, in the hundreds or thousands,
  correspond to log-likelihood gains that far exceed the $+1$ cost of adding $q$. Together, $R^2$ tells us how well the
  Mandelbrot family fits; KS tells us whether the fit breaks down in a specific rank region; and AIC tells us whether
  the extra parameter $q$ is worth its penalty.

 \subsection{Results}
  \label{subsec:results}

  \begin{table}[h]
  \caption{Shared Mandelbrot fit quality by model and domain. Each cell reports the $R^2$ of an independently estimated
  two-parameter Mandelbrot fit. The practical reading is that the rank-frequency signal is stable enough across the
  measured cohort to justify a common reference baseline.}
  \label{tab:fit-quality}
  \centering
  \small
  \setlength{\tabcolsep}{4pt}
  \begin{tabular}{lcccccc}
  \toprule
  Domain & GPT-5.1 & Claude Sonnet 4.6 & Llama 3.1 8B & Gemini 2.5 Pro & Mistral Large & Qwen 2.5 7B \\
  \midrule
  Global       & 0.956 & 0.961 & 0.967 & 0.966 & 0.972 & 0.967 \\
  News         & 0.960 & 0.952 & 0.946 & 0.930 & 0.909 & 0.950 \\
  Biomedical   & 0.971 & 0.963 & 0.967 & 0.971 & 0.970 & 0.967 \\
  Legal        & 0.972 & 0.959 & 0.968 & 0.966 & 0.960 & 0.966 \\
  Code         & 0.957 & 0.976 & 0.971 & 0.963 & 0.974 & 0.969 \\
  Social media & 0.961 & 0.950 & 0.948 & 0.959 & 0.943 & 0.949 \\
  \bottomrule
  \end{tabular}
  \end{table}

  Every $R^2$ in Table~\ref{tab:fit-quality} exceeds 0.90; 34 of 36 exceed 0.94. The Mandelbrot form is not merely
  adequate; it is nearly saturated across the full six-model cohort. The residual variance (the gap between observed
  rank frequency and the two-parameter fit) is dominated by the head of the distribution (the top 10 ranks), where the
  coding theoretical cost argument is expected to break down because Huffman-style prefix codes further optimize the top
   tokens. Notably, Qwen 2.5 7B, the smallest model in the cohort, fits as tightly as the frontier models: convergence
  is not an artifact of scale. We report $R^2$ as a readable summary of fit quality, but the load-bearing tests of the
  Mandelbrot form against alternatives are the AIC comparisons (Section~\ref{subsec:results}, Model selection) and the
  KS statistics on the empirical CDF, not the log-log $R^2$ (cf.\ Clauset, Shalizi, and Newman [33] on the standard
  cautions for power-law fits).

  \textbf{Model selection.} In 35 of the 36 model-by-domain pairs we tested, AIC prefers the two-parameter Mandelbrot
  form over the one-parameter Zipf form, with $\Delta$AIC values ranging from $+18$ to $+15{,}094$. The single tie,
  Claude on legal text, has a $\Delta$AIC of $-1$, which is within the noise of the comparison and occurs precisely at
  the corner where the fitted $q$ approaches zero and the Mandelbrot form collapses to Zipf. The fitted exponents $s$
  range from 0.88 to 1.11; the fitted shift parameters $q$ range from 0.04 to 5.81. No model or domain produces a fit
  that escapes the Mandelbrot family.

\begin{figure}[htbp]
      \centering
      \includegraphics[width=\textwidth]{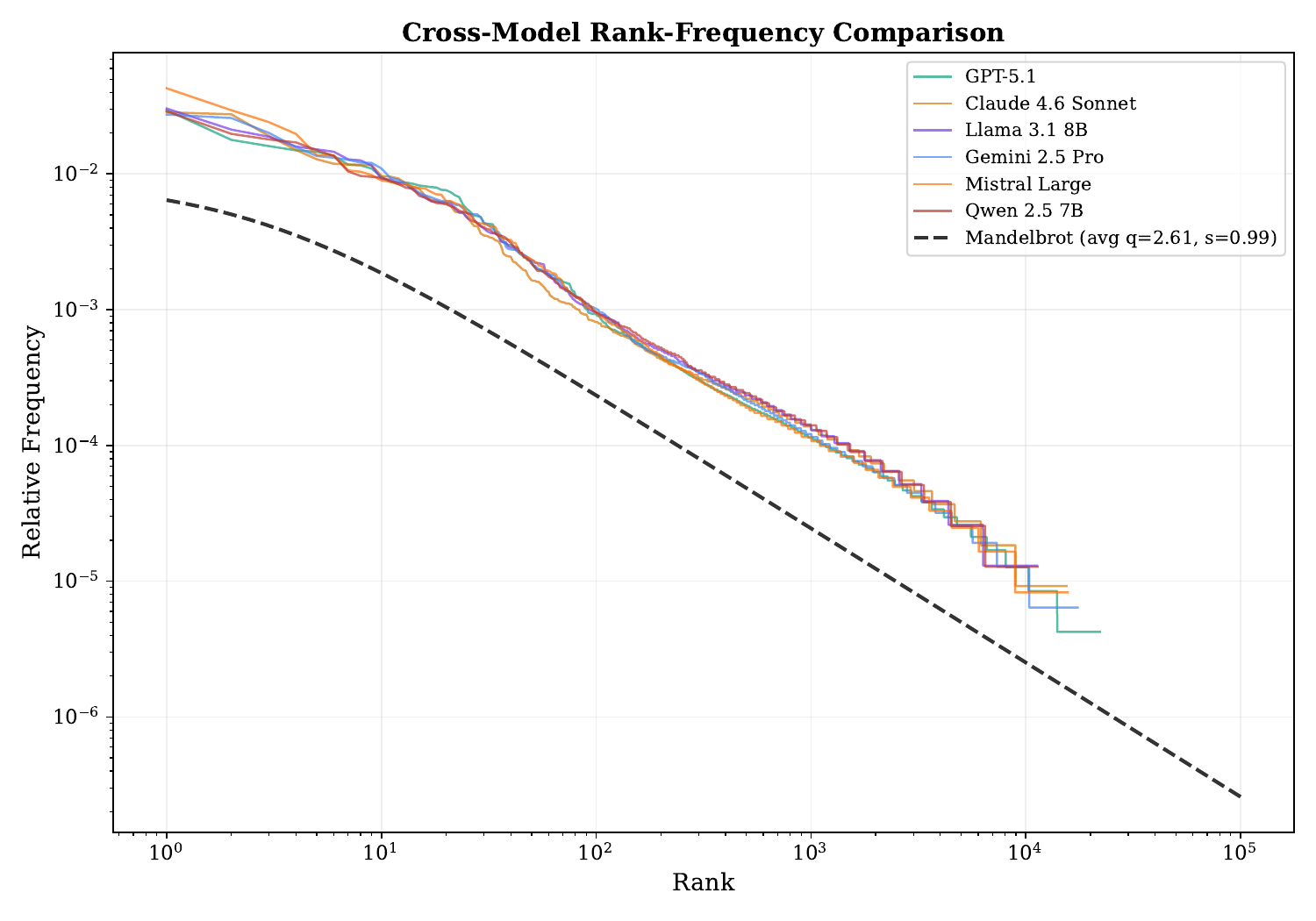}
      \caption{Cross-model overlay of normalized rank-frequency curves. Across the distribution, vendor differences are
  smaller than domain differences. This is the regime in which a shared first-pass baseline becomes operationally
  useful.}
      \label{fig:cross-model-overlay}
  \end{figure}

  \textbf{Cross-model overlay.} When normalized rank-frequency curves for all six models are plotted on the same axes,
  they closely overlap (Figure~\ref{fig:cross-model-overlay}). The scatter between GPT-5.1, Claude Sonnet 4.6, Llama 3.1
   8B, Gemini 2.5 Pro, Mistral Large, and Qwen 2.5 7B in the body of the distribution (ranks 10 to 10{,}000, which carry
   almost all the information content) is smaller than the scatter between domains within a single model. Put
  differently, a GPT-5.1 news output looks more like a Qwen 2.5 7B news output than it looks like a GPT-5.1 code output.
   The domain identity, not the model identity, is what moves the curve.

  \begin{figure}[htbp]
      \centering
      \includegraphics[width=\textwidth]{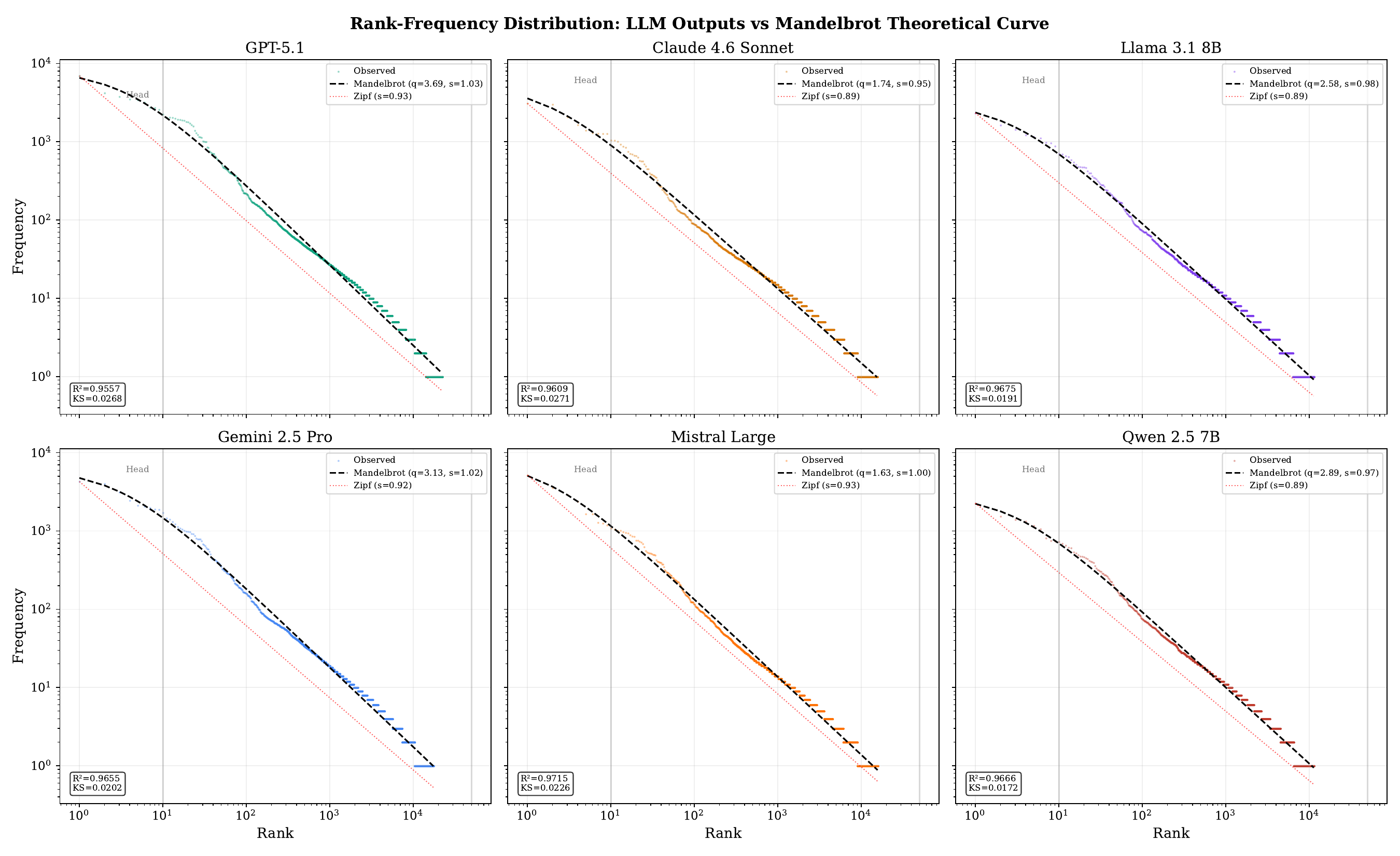}
      \caption{Per-model Mandelbrot fits. The Mandelbrot form consistently captures the head flattening that the simpler
   Zipf form misses, which is why the extra parameter earns its keep in the measured cohort.}
      \label{fig:rank-frequency-curves}
  \end{figure}

  \begin{figure}[htbp]
      \centering
      \includegraphics[width=\textwidth]{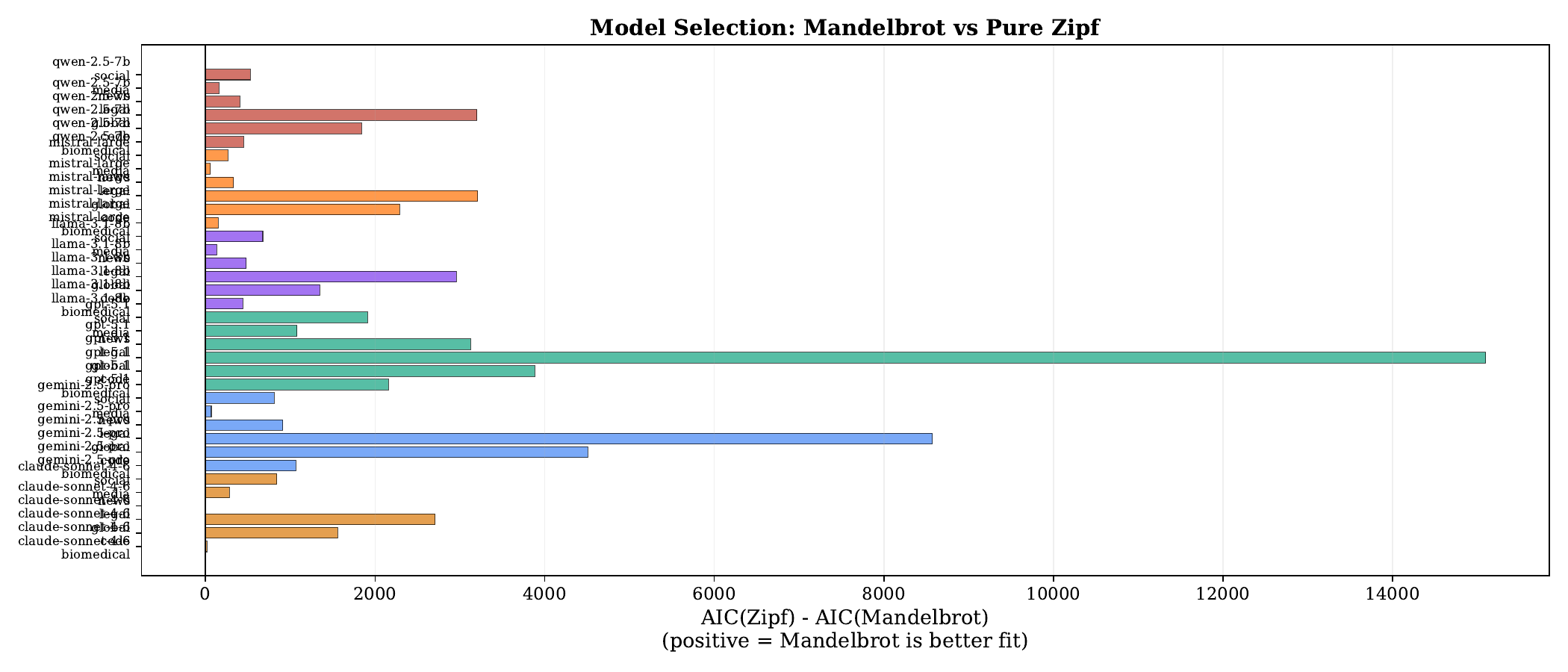}
      \caption{AIC model selection. Positive $\Delta$AIC values favor Mandelbrot over Zipf. The practical point is not
  aesthetic fit, but model selection: the richer form is repeatedly justified by the data.}
      \label{fig:aic-comparison}
  \end{figure}

  \begin{figure}[htbp]
      \centering
      \includegraphics[width=\textwidth]{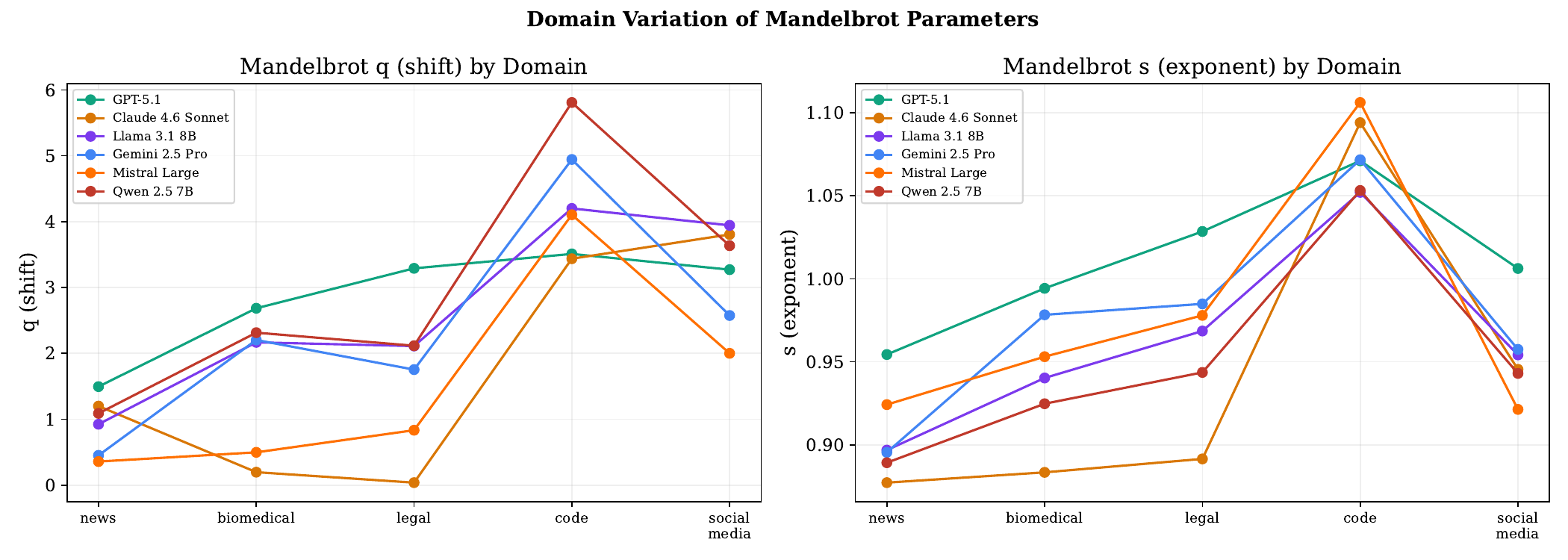}
      \caption{Domain parameter variation. Within-model dispersion across domains exceeds between-model dispersion at a
  fixed domain, reinforcing the systems point that domain matters more than vendor on this axis.}
      \label{fig:domain-variation}
  \end{figure}

\subsection{What the Convergence Does and Does Not Show}

  The convergence does not show that all LLMs are the same model. They manifestly are not: they differ in capability,
  style, refusal behavior, context length, and countless downstream benchmarks. What the convergence does show is a
  two-part structural result. On the token rank-frequency axis, all six models converge to the Mandelbrot family. Two
  pieces of prior theory make the family's relevance plausible at the outset. Cross-entropy training on corpora whose
  own rank-frequency statistics follow the Mandelbrot form preserves that form under regularity conditions. Cugini et
  al.\ [26] establish that i.i.d.\ ranked samples admit a local Zipf-Mandelbrot approximation within a rank-neighborhood
   validity window (of width $O(N^{5/6})$) that covers our fitted range. Neither argument directly predicts what our
  data show: LLM outputs are autoregressive rather than i.i.d., yet they conform; our single two-parameter fit holds
  globally across more than three decades of rank rather than locally at a single position with different exponents at
  different positions; and the fitted $(q, s)$ pair is recovered across six independently-trained models with no
  a-priori reason to share a parametric neighborhood. The $R^2 > 0.94$ in 34 of 36 conditions, the adequacy of two
  parameters rather than three or four, and the cross-vendor parameter stability together constitute the empirical
  content of this level of the result. The second part is what the universality result does not predict. Within that
  shared family, different training data, alignment procedures, and pretraining pipelines produce consistently different
   parameter values. The family is inherited from the objective; the specific location within the family is not. The
  fingerprinting analysis in Section~4.5 quantifies this second part directly: cross-model parameter spread exceeds
  within-model bootstrap noise by more than an order of magnitude. That the shared objective pins down the family but
  not the parameters is the structural result that makes both a common baseline and statistical fingerprinting
  simultaneously possible.

  A theoretical point deserves careful attention here. Cugini et al.\ [26] prove that any sufficiently large ranked
  sample drawn i.i.d.\ from an analytic parent distribution admits, in the vicinity of any fixed rank position
  $\lambda_0 = r_0/N$, a local Zipf-Mandelbrot approximation with error $O((\lambda - \lambda_0)^3)$, statistically
  indistinguishable from the true order statistics over a rank-neighborhood window of width $O(N^{5/6})$. For vocabulary
   sizes characteristic of modern BPE tokenizers ($N$ of order $10^5$), this window is of order $10^4$ ranks wide; our
  fits, which span ranks 10 to $10^4$, lie inside it. The existence of some Zipf-Mandelbrot-like fit is therefore
  broadly consistent with the theorem, and we use the theorem as a prior on our choice of functional family rather than
  as a proof of our empirical finding. Three aspects of our findings, however, are not consequences of it. First, the
  theorem assumes i.i.d.\ sampling; LLM outputs are autoregressive and context-conditioned, and the observation that
  they nevertheless conform to the same family is an empirical regularity that the theorem does not predict. Second,
  Cugini's result describes a local approximation pinned to a specific $\lambda_0$, with a potentially different $(A, B,
   \alpha)$ triple in different neighborhoods (the paper's Fig.\ S1 illustrates exactly this for a Gaussian parent,
  where different local exponents arise at different positions within the same sample). Our data fit a single
  two-parameter $(q, s)$ globally across more than three decades of rank with $R^2 > 0.94$, which is a stronger
  empirical property than the local theorem guarantees. Third, even granting that some Zipf-Mandelbrot-like fit is
  expected per sample, nothing in the theorem predicts that six independently trained models with distinct training
  data, alignment procedures, and pretraining pipelines should land in a narrow shared parametric region. The
  universality result and our empirical result are therefore parallel and complementary rather than duplicative: Cugini
  provides a strong prior on fit quality for any one sample; our data add that the prior is fulfilled under non-i.i.d.\
  generation, with a global rather than piecewise fit, and with cross-vendor parameter stability. Separately, and this
  remains valid under either reading of the theorem's scope: precisely because the Mandelbrot form is an expected
  attractor under broad conditions, departures from a cohort-level baseline become interpretable as genuine anomalies
  rather than artifacts of an arbitrary reference choice.

  Three consequences follow.

  First, the Mandelbrot distribution should be treated here as a strong empirical regularity on the token rank-frequency
   axis, not as a claim that all language models obey a universal law in every regime. Wang et al.\ [27] have
  independently documented that LLM-generated text obeys multiple linguistic scaling laws, including Zipf and Heaps,
  with exponents close to those of human text, confirming that the regularity reported here is not an artifact of our
  particular experimental design. Notably, their treatment of what they term ``Mandelbrot's law'' refers to multifractal
   detrended fluctuation analysis of text embeddings, which is a distinct mathematical object from the classical
  Mandelbrot rank-frequency distribution used in this paper. Within the measured cohort, the regularity is stable enough
   to justify a shared baseline; whether it extends unchanged to small models, base models, multilingual systems, or
  future synthetic-data-heavy training pipelines remains an empirical question. Concurrent work by Zhu et al.\ [34]
  applies the Zipf-Mandelbrot law to AI-generated content across 15 human-corpus comparison baselines and reports that
  fitted parameters vary across large models; that work characterizes AIGC distributional properties for
  information-resources management at the word level, whereas our contribution operates at the BPE-token level, derives
  a scoring primitive from a variational posterior, and frames the parameter variation as statistical fingerprinting for
   provenance verification. Token-level peer-reviewed evidence adjacent to the present finding has appeared in parallel:
   Mikhaylovskiy [36] reports that Zipf's and Heaps' laws hold for BPE-tokenized LLM outputs across eleven open-weight
  models in four families (Qwen, Llama, Granite, Mistral) but only within a narrow, model-dependent sampling-temperature
   range, interpreted through a phase-transition framework. Our result is complementary along three axes: we use the
  two-parameter Mandelbrot form rather than a single-exponent Zipf and obtain a global fit across more than three
  decades of rank ($R^2 > 0.94$ in 34 of 36 conditions); we observe convergence at each vendor's default decoding rather
   than only at a tuned critical temperature; and we use the regularity as the basis for a scoring primitive and a
  fingerprinting pipeline rather than as a phase-transition diagnostic.

  Second, regularity is still operationally valuable. A single rank table, computed once from a reference corpus, can be
   used to score outputs from every model in the measured cohort and plausibly from nearby frontier systems, even though
   those models differ sharply in capability, style, context length, and alignment behavior.

  Third, deviations from the shared signature are a candidate diagnostic signal. Because the baseline is stable on the
  measured axis, statistically significant departures can serve as alerts for fine-tuning, damage, production drift, or
  unusual decoding regimes. The experiments in this paper motivate that possibility; they do not yet fully validate each
   downstream diagnosis.

  A broader universality question, which this paper does not directly answer but against which the present finding
  should be positioned, is whether frontier LLMs converge beyond the token rank-frequency axis. Huh et al.\ [37] have
  argued for a Platonic Representation Hypothesis: that neural networks trained on different data, architectures, and
  objectives converge toward a shared representation of reality, measured at the level of internal embedding kernels via
   centered kernel alignment and pointwise mutual information. The convergence reported here is compatible with that
  hypothesis but operates on a distinct, more API-accessible object (the token rank-frequency distribution of model
  outputs) rather than on internal representations that require weight access. The two findings are complementary:
  shared training distributions produce shared internal-representation structure in their framing and, as this paper
  adds, shared output-distribution structure within which per-vendor fingerprints remain cleanly recoverable (C2).

  \subsection{How Robust Is Convergence}

  We anticipate four concerns about the convergence claim and address them briefly here.

  \textbf{Is it a tokenizer artifact?} The reported convergence is measured under a deliberate methodological choice:
  all six models' outputs are re-tokenized with the Llama 3.1 8B BPE vocabulary, so rank-frequency distributions are
  compared in a shared token space. This establishes the operational claim the paper actually depends on: for any
  black-box output, an auditor controls the tokenizer used to compute ranks, so the relevant question is whether a
  single chosen tokenization yields a stable baseline across vendors. Within the measured cohort, it does. It does not
  establish the stronger, tokenizer-invariance claim that each vendor's native tokenizer would recover identical fitted
  $(q, s)$ without this normalization step; that stronger claim would require re-running the experiment under each
  model's native vocabulary and comparing fitted parameters directly, which we scope as roadmap work (Section~8.1). The
  result we report should therefore be read as: once outputs are mapped into a common token space, the Mandelbrot family
   fits uniformly well across vendors, and the parameter separability that drives the fingerprinting result
  (Section~4.5) is preserved under a single, fixed normalization.

  \textbf{Is it a training data artifact?} Partly, yes, and we take this seriously rather than defend against it. The
  six models were trained on overlapping web-scale corpora, so complete statistical independence of training sources is
  not part of the claim. Two considerations keep the result substantive nonetheless. First, the cross-entropy training
  objective (Section~3.3) predicts some form of rank-frequency regularity in outputs whenever training corpora
  themselves exhibit one, under conditions weaker than independence of training data, and Cugini et al.\ [26] establish
  a complementary universality prior for i.i.d.\ ranked samples. These theoretical backdrops make the shared baseline
  the auditor relies on plausible without requiring independent rediscovery of the law; what the data establish is that
  the baseline is stable empirically across vendors, domains, and scales, and sharp enough globally (rather than
  piecewise) to be useful as a reference. Second, and more importantly, the parameter-level fingerprinting result
  (Section~4.5, Table~\ref{tab:fingerprint}) is not explained by shared training data. If the six models were statistical duplicates due to
   overlapping corpora, their fitted $(q, s)$ values should cluster tightly. They do not. The cross-model $q$ spread
  (1.63 to 3.69) exceeds within-model bootstrap standard deviations (0.03 to 0.10) by more than an order of magnitude,
  meaning that whatever each vendor does downstream of the shared corpus (tokenization choices, architecture scale,
  alignment procedures, decoding defaults) moves them to consistently different points in the parameter space. The
  substantive finding is therefore not that shared data produces shared statistics; it is that shared data plus distinct
   training pipelines produces a shared family with distinguishable fingerprints.

  \textbf{Does RLHF break it?} Empirically, no. All six of GPT-5.1, Claude Sonnet 4.6, Llama 3.1 8B, Gemini 2.5 Pro,
  Mistral Large, and Qwen 2.5 7B are post-RLHF, post-Constitutional AI, or post-equivalent instruction tuning; their
  outputs still fit Mandelbrot with $R^2$ above 0.90. Alignment procedures concentrate probability mass on the ``helpful
   and harmless'' region of response space but do not appear to reshape the rank frequency law.

  \textbf{Does aggressive quantization break it?} Empirically, no. Llama 3.1 8B was served locally via Ollama using
  Q4\_K\_M 4-bit quantization (see Section~3.1) and still fits Mandelbrot across all six conditions (the five
  single-domain tables plus the global aggregate), with $R^2$ above 0.94. Quantization to 4 bits is a substantial
  perturbation of the logits, sufficient to shift many benchmark scores, but it does not shift the rank-frequency law.
  Whether more aggressive schemes (2-bit quantization, pruning, distillation) preserve the law is open to debate.

  \textbf{Do smaller or older models break it?} We are extending the study along this axis. Pre-trained only base models
   and small ($<$ 1B parameter) models are candidates where the fit could degrade. An $R^2$ that falls materially below
  0.91 would itself become an assessment signal, flagging the model as distributionally anomalous.

  Finally, a note on parameter identifiability. Manin [28] observes that under Mandelbrot's original cost-optimization
  derivation, $q$ and $s$ are not fully independent: they co-vary along the optimization surface. In practice, this
  means that the two-dimensional fingerprint space (Table~\ref{tab:fingerprint}) carries somewhat less information than two unconstrained
  degrees of freedom would suggest. Our bootstrap confidence intervals already reflect this coupling empirically, but we
   flag the theoretical constraint for readers who may otherwise treat $q$ and $s$ as orthogonal.

  \section{What the Shared Signature Makes Possible}
  \label{sec:applications}

  Because the Mandelbrot baseline is model-agnostic within the measured cohort, stable across frontier models on the
  rank-frequency axis, and computable in $O(1)$ per token, it can be composed into assessment pipelines in ways that are
   impossible for more expensive or model-specific signals. The key practical consequence is not philosophical
  convergence; it is the availability of a cheap scoring layer that can run on a CPU at scale from black-box outputs.
  The following eight applications are direct or near-term consequences.

  \subsection{A Model-Agnostic Grounding Score for Any LLM Output}
  \label{subsec:grounding-score}

  For any generated passage, rank deviation (Eq.~\eqref{eq:rankdev}, taken with respect to the reference rank table
  introduced in Section~2.2: the global Wikipedia rank table in this paper) and log-space deviation
  (Eq.~\eqref{eq:logratio}) produce per-token scores that depend only on the output text and the public rank table. No
  model access, no API calls, no white-box information is required. The same scoring code runs against GPT outputs,
  Claude outputs, Llama outputs, and future-model outputs without modification. This gives auditors and third-party
  evaluators a first-class signal they can compute at essentially zero cost per example. Section~5 instantiates this
  application as a concrete scoring primitive, reports pilot performance across three public benchmarks, and explicitly
  notes where the primitive succeeds and where it cannot succeed structurally.

  \subsection{Training Health Monitoring}

  During pre-training and fine-tuning, the rank-frequency distribution of model outputs can be continuously monitored.
  Fitted $(C, q, s)$ values that drift away from the reference values on held-out prompts indicate that the model's
  output distribution is departing from the Mandelbrot reference. Three candidate failure modes plausibly produce
  distinctive signatures (mode collapse via tail flattening, overfitting on a narrow corpus via $q$ shift, catastrophic
  forgetting after fine-tuning via discontinuous parameter jumps). Mapping specific failure modes to specific parameter
  movements is a hypothesis, not a measured result; it would need to be established on held-out training runs before
  being relied on operationally.

  We propose Mandelbrot fit deviation as a canary metric for training runs, tracked alongside loss and evaluation
  accuracy. Unlike loss, which is computed against the training distribution, the Mandelbrot fit is computed against an
  external reference and therefore detects problems that loss, by construction, cannot see.

  \subsection{Fine-Tuning Damage Detection}

  A large fraction of production LLM incidents comes from fine-tuning that silently degrades the base model. Classical
  evaluation suites detect damage only if it falls within an already curated benchmark. Mandelbrot fit change is a
  complementary, unconditional signal. Given a base model and a fine-tuned variant, the difference between their fitted
  $(q, s)$ on a common prompt set quantifies how much the fine-tuning has reshaped the output distribution.
  Conceptually, this uses the base model itself as a local reference (a pre-alignment distribution per prompt set),
  which is strictly more sensitive than comparing to the shared Mandelbrot baseline because small RLHF or SFT-scale
  shifts can be statistically significant against the base while still fitting Mandelbrot at the global level. This is
  cheap, reproducible, and independent of any particular evaluation benchmark. A fine-tune that preserves capabilities
  on curated benchmarks but shifts Mandelbrot parameters outside their normal range is a candidate for further scrutiny.

  \subsection{Production Distribution Shift and Jailbreak Detection}

  In deployed systems, outputs can drift for reasons external to the model: changes in the input prompt distribution,
  prompt injection attacks, jailbreak prompts that push the model into undertrained regions of response space, or
  infrastructure problems that corrupt decoding (for example, temperature set incorrectly, sampler misconfiguration, KV
  cache reuse bugs). All four of these failure modes perturb the rank-frequency profile of outputs in ways that a
  Mandelbrot monitor can detect. A lightweight streaming estimator of $(q, s)$ over a rolling window of the last $N$
  outputs, compared to the reference fit, flags anomalous periods for investigation. This is orders of magnitude cheaper
   than running a dedicated hallucination detector on every output.

  \subsection{Statistical Model Fingerprinting and Output Provenance}
  \label{sec:fingerprinting}

  Although all six models we tested sit inside the same Mandelbrot family, they do not collapse to identical parameter
  values. Global $(q, s)$ estimates span a usable range (Table~\ref{tab:fingerprint}). That means the shared signature
  is useful in two complementary ways: it supports a common baseline for black-box scoring, and it preserves enough
  model-specific nuance to support statistical fingerprinting and provenance analysis.

  \begin{table}[h]
  \caption{Global fingerprint parameters by model. The table illustrates the paper's key claim of provenance: frontier
  models can share the same low-dimensional family while remaining statistically distinguishable through the nuances of
  their fitted parameters.}
  \label{tab:fingerprint}
  \centering
  \small
  \begin{tabular}{lcccc}
  \toprule
  \textbf{Model} & $q$ & $s$ & $C$ & $R^2$ \\
  \midrule
  GPT-5.1            & 3.69 & 1.03 & 32{,}082 & 0.956 \\
  Gemini 2.5 Pro     & 3.13 & 1.02 & 20{,}110 & 0.966 \\
  Qwen 2.5 7B        & 2.89 & 0.97 & 8{,}352  & 0.967 \\
  Llama 3.1 8B       & 2.58 & 0.98 & 8{,}215  & 0.968 \\
  Claude Sonnet 4.6  & 1.74 & 0.95 & 9{,}355  & 0.961 \\
  Mistral Large      & 1.63 & 1.00 & 13{,}312 & 0.972 \\
  \bottomrule
  \end{tabular}
  \end{table}

  The cross-model spread in $q$ (1.63 to 3.69) exceeds its per-model bootstrap standard deviation (typically 0.03 to
  0.10) by more than an order of magnitude. Framed as a signal-to-noise ratio, this corresponds to tens of standard
  deviations of separation between model families once a few thousand output tokens are available for fitting. This is
  the empirical foundation for a lightweight fingerprinting pipeline in which a held-out auditor can estimate which
  model most likely generated a body of text, detect silent model substitution in production, or test whether a
  vendor-delivered output is statistically consistent with the model family it purports to belong to. The method is not
  a cryptographic guarantee, but it is a practical provenance signal that can complement watermarking and operational
  monitoring.

  \begin{itemize}
      \item \textbf{Provenance verification.} Confirming that a vendor has delivered outputs from the model they claim.
      \item \textbf{Pipeline integrity.} Detecting cases where a downstream component has silently substituted a cheaper
   model.
      \item \textbf{Watermark complement.} Providing a weak but model-agnostic provenance signal even when cryptographic
   watermarks are absent or stripped.
      \item \textbf{Wrapper and arbitrage detection.} Commercial services that claim to serve a premium model but route
  requests to a cheaper model to capture the price differential leave a statistical trace in the fingerprinting space.
  The same apparatus flags silent model substitution by upstream aggregators.
  \end{itemize}

  The approach is not a cryptographic guarantee. It is a statistical test with quantifiable false-positive and
  false-negative rates.

  We note that recent work has independently demonstrated that distributional signatures function as passive model
  fingerprints. McGovern et al.\ [29] show that part-of-speech frequency distributions, constituency parse frequencies,
  and top-$k$ token histograms distinguish model families at the COLING 2025 GenAIDetect workshop, and Fu et al.\ [30]
  achieve strong fingerprinting accuracy using learned features over bilingual corpora. Our contribution differs in two
  respects: we derive the fingerprint from a theoretically motivated rank-frequency law rather than from learned or
  POS-level features, and we use the same distributional machinery for both scoring (Section~5) and provenance (this
  section). An important methodological caution follows: because the scoring primitive and the fingerprint share a
  common statistical substrate, any deployment must verify that the score varies with output quality, holding the model
  fixed, not merely with model identity content fixed. The benchmark results in Section~5 provide this verification for
  the scoring application, since all pilot evaluations are conducted within a single model (Llama 3.1 8B).

  \subsection{Synthetic Text Auditing}

  Human text also follows the Mandelbrot law. Frontier LLM outputs produce slightly steeper fits than their training
  corpora, because instruction tuning and RLHF concentrate probability mass in ways that empirical distributional
  measurements have confirmed. Shumailov et al.\ [12] demonstrate that LLM outputs lose tail mass relative to their
  training distribution, and Kirk et al.\ [13] quantify how RLHF reduces output diversity. The difference is small per
  passage but aggregates cleanly over paragraphs. Mandelbrot fit residuals become one feature (among several) in
  synthetic text classifiers, with the specific advantage that the feature is theoretically grounded rather than a
  learned artifact of a particular detector model. This is particularly useful in adversarial settings where a
  paraphrase attack might defeat a learned classifier but cannot easily rewrite a passage to match both the paraphrase
  semantics and the target Mandelbrot parameters.

  \subsection{Cross-Model Consensus Scoring}

  The shared Mandelbrot baseline provides a way to score factuality without relying on any individual model as the
  ground truth. Given a prompt, generate outputs from $N$ independent models. Compute rank deviation $\Delta_r$ for each
   named entity or quantitative claim in each output. A claim with consistently low rank deviation across models is
  either a common truth (distributionally central) or a common failure mode. A claim with consistent rank deviation
  patterns across models that are atypical for the local context is a candidate for flagging. Importantly, because the
  baseline is the same across models, the cross-model comparison is strict apples-to-apples. This is harder to do with
  model-specific uncertainty estimators, whose scales are not comparable.

  \subsection{Benchmark Contamination Diagnostics}

  Leakage of evaluation corpora into training data is a recurring problem. One symptom is that a model's output on
  contaminated prompts is unusually close to a specific reference passage. Against the global Mandelbrot baseline, such
  ``echo'' outputs exhibit a characteristic rank deviation profile, with specific low-frequency tokens from the
  memorized passage appearing at anomalously high local ranks. The signature is subtle on any one output but accumulates
   across a benchmark. Running Mandelbrot-based contamination audits is cheap and does not require access to the
  training data, which is usually the bottleneck for contamination studies. We flag this as a specialized diagnostic,
  not a general-purpose one: most production assessment pipelines will never need it, but the audit teams that do will
  find it cheap to deploy.

  \subsection{Summary of Application Landscape}

  The eight applications above differ sharply in maturity, and the reader should calibrate expectations accordingly. The
   central validated contribution of this paper is the scoring primitive developed in Section~5, supported by pilot
  results across three public benchmarks. Statistical model fingerprinting (Section~\ref{sec:fingerprinting}) is the
  next most concrete, because it follows directly from the measured spread in fitted parameters, even though a full
  operating-characteristic study remains future work. The remaining applications (training health monitoring,
  fine-tuning damage detection, production shift detection, synthetic text auditing, cross-model consensus scoring, and
  benchmark contamination diagnostics) are enabled directions rather than validated products. We include them because
  they illustrate the range of problems that a cheap, model-agnostic distributional baseline can address in principle,
  but we caution that each requires its own empirical validation before operational deployment.

\section{A Concrete Scoring Primitive for Black-Box LLM Assessment}
  \label{sec:primitive}

  The rest of the paper develops one application in sufficient detail to make the value proposition concrete: a low-cost
   scoring primitive designed for black-box LLM and agent outputs, intended to run on CPU and to compose with heavier
  verification tools rather than replace them.

  \subsection{Design}
  \label{sec:primitive-design}

  For each token in the generated output, compute the following quantities in hybrid mode:

  \begin{itemize}
      \item the LLM log probability $\log P_{\text{LLM}}(t \mid \text{context})$, obtained from the model's logprobs API
   or from a local forward pass.
      \item the Mandelbrot log probability $\log P_{\text{RI}}(t)$, obtained from the pre-built rank table.
      \item the log-space deviation $\delta_{\log}(t \mid c)$ from Eq.~\eqref{eq:logratio}.
      \item a posterior weighted score from Eq.~\eqref{eq:posterior} with a domain-specific $\beta$ from
  Eq.~\eqref{eq:beta}.
  \end{itemize}

  Aggregate token-level scores to larger units by weighted mean, max, or threshold proportion. The aggregation unit
  varies by use case: in this paper, a span is a contiguous token range corresponding to an annotated region in a
  benchmark; an entity is a named entity extracted by a standard NER pipeline; and a full response is the entire
  generated passage. In practice, the primitive has two deployable modes. The hybrid mode uses model logprobs when
  available. The pure rank mode uses only tokenization, lightweight entity extraction, and the reference rank table,
  making it usable for closed APIs and high-volume agent traces where only the output text is observable.

  \subsection{Pilot Validation on Public Benchmarks}
  \label{sec:primitive-pilot}

  We ran the primitive as a scoring component on three public hallucination and faithfulness benchmarks, using Llama 3.1
   8B as the scoring model, a rank table built from a 4-billion-token Wikipedia snapshot, and $\beta$ fixed at 1 as a
  calibration baseline.

  \textbf{FRANK (summarization faithfulness) [14].} FRANK provides 6{,}356 human-annotated error spans across 2{,}246
  summaries of CNN/DailyMail and XSum articles, with error types labeled according to a seven-category schema. Under the
   benchmark-level protocol used for the headline pilot result, the log-space primitive against the global Wikipedia
  rank table yields a span-level ROC-AUC of 0.585. The important result is the breakdown by error type: out-of-article
  entities (AUC 0.646, Kolmogorov-Smirnov $p < 10^{-10}$) and entity substitutions (AUC 0.562, $p < 10^{-10}$) produce
  strong separations, while coreference and discourse-link errors do not ($p > 0.2$). The rank correlation between
  annotated error type and AUC is $\rho = +0.81$ ($p = 0.05$), indicating a monotonic decrease in AUC with
  distributional error type. That per-type pattern, not the aggregate 0.585, is what supports the taxonomy developed in
  Section~\ref{sec:taxonomy}.

  \textbf{TruthfulQA [15] (817 knowledge-grounded multiple-choice questions over 38 categories).} Using the scoring
  primitive (global Wikipedia baseline, as with FRANK) to select among candidate answers improves multiple-choice
  accuracy from 61.4\% to 63.8\%. The category-level picture is more informative than the aggregate: temporal indexical
  errors achieve AUC 0.940, locational indexical errors 0.821, advertising 0.757, and subjective claims 0.806.
  Categories whose errors are expressed in domain-appropriate vocabulary (science, politics, psychology, identity
  indexical) have AUCs of 0.42 to 0.45, clearly below chance in some cases. The category-level bifurcation, strong on
  surface distributional anomalies and weak on reasoning errors in normal vocabulary, is consistent with the taxonomy
  introduced next.

  \textbf{HaluEval [16] (30{,}000 examples across QA, dialogue, and summarization).} We use the global Wikipedia
  baseline throughout HaluEval; the primitive never reads the source article, so the signal cannot be repackaged ROUGE
  (a concern we return to in Section~\ref{sec:related}). The QA subset is dominated by a length artifact (length alone
  achieves AUC 0.965), leaving the primitive with no additional signal to contribute. On the dialogue subset, where
  length alone reaches only AUC 0.703, the primitive contributes a small but statistically strong additional signal
  ($+0.017$ AUC beyond length, Mann-Whitney $p < 10^{-69}$). On the summarization subset, the primary error type is
  paraphrastic, and the primitive is at chance level.

  These results position the primitive not as a replacement for sampling-based detectors, but as a fast component in
  compound evaluation systems. Its strongest value is that it supplies cheap evidence exactly where a distributional
  signal should help: lexical anomalies, unsupported surface-form entities, and other rank-based deviations. That value
  increases further when the rank-only black-box mode can be used without logprobs.

   \subsubsection{Three Token-Filter Aggregations at the Span Level}
  \label{sec:primitive-aggregations}

  The headline results in Section~\ref{sec:primitive-pilot} use benchmark-specific protocols chosen for each dataset.
  Table~\ref{tab:span-aggregations} serves a different purpose: it compares three aggregation schemes under a consistent
   internal setup, so the practical contribution of entity filtering and rank-only scoring can be seen directly. The
  numbers in Table~\ref{tab:span-aggregations}, therefore, should not be read as a verbatim restatement of the
  Section~\ref{sec:primitive-pilot} headline AUCs.

  \begin{itemize}
      \item \textbf{Output-level.} Mean $\delta_{\log}$ over every token. Requires logprobs. This is the legacy
  whole-output baseline used here for within-table comparison, but it is not identical to the benchmark-specific
  headline protocol reported in Section~\ref{sec:primitive-pilot}.
      \item \textbf{Span with entity-token filter.} The same quantity is restricted to the NER-tagged token positions
  (persons, organizations, locations, dates, numeric quantities) from a standard spaCy pipeline that falls inside the
  span. Filters the within-span signal to select the tokens most likely to carry factual risk, while leaving the span
  itself as the classification unit.
      \item \textbf{Rank-only at entity tokens.} At the same entity positions, with no logprob term. Requires no
  logprobs at all, which makes it applicable to closed APIs (Anthropic, some Gemini endpoints) that do not expose
  token-level probabilities.
  \end{itemize}

  Table~\ref{tab:span-aggregations} reports the three span-level aggregates under a common comparison setup at $\beta =
  1$.

  \begin{table}[htbp]
  \centering
  \caption{Span-level AUCs under a common comparison setup (llama-3.1-8b, $\beta = 1$). The important systems result is
  that the entity-filtered and rank-only variants outperform the output-level aggregate on the distributional slices,
  making the black-box mode operationally credible.}
  \label{tab:span-aggregations}
  \begin{tabular}{lccc}
  \toprule
  \textbf{Benchmark (slice)} & \textbf{Output-level} & \textbf{Span + entity-token filter} & \textbf{Rank-only} \\
  \midrule
  FRANK overall                & 0.466 & 0.558          & 0.559          \\
  FRANK RelE (relation)        & 0.498 & 0.594          & \textbf{0.604} \\
  FRANK EntE (entity)          & 0.531 & \textbf{0.594} & 0.592          \\
  FRANK LinkE (link)           & 0.512 & 0.572          & \textbf{0.579} \\
  FRANK OutE (out-of-article)  & 0.430 & 0.558          & \textbf{0.559} \\
  FRANK CorefE (coreference)   & 0.407 & 0.461          & 0.468          \\
  FRANK CircE (circumstantial) & 0.382 & 0.490          & 0.491          \\
  HaluEval summarization       & 0.321 & 0.514          & 0.514          \\
  HaluEval dialogue            & 0.478 & 0.494          & 0.507          \\
  HaluEval qa                  & 0.158 & 0.385          & 0.457          \\
  TruthfulQA overall           & 0.513 & 0.509          & 0.521          \\
  \bottomrule
  \end{tabular}
  \end{table}

  Two findings follow.

  First, the span with the entity-token filter consistently outperforms the output-level aggregate on every slice where
  the expected error type is distributional. The largest lifts occur on FRANK out-of-article entities ($+0.128$ over
  output-level), HaluEval summarization ($+0.193$), and FRANK circumstantial errors ($+0.108$). This empirically
  confirms the Section~\ref{sec:taxonomy} taxonomy: filtering to the tokens most likely to carry factual content
  concentrates the within-span signal that the primitive measures.

  Second, \textbf{the rank-only aggregate matches or exceeds the span with entity-token filter everywhere} (within $\pm
  0.01$ on every FRANK slice and across HaluEval and TruthfulQA). The rank-only signal requires no logprobs. This is the
   black-box result: the paper's framing in Section~\ref{sec:applications} has been promising: for distributional error
  types, the primitive delivers an error-detection signal comparable to the logprob-using variant, using only the
  tokenization and the reference rank table, from an API that exposes neither internal probabilities nor the generation
  sampler.

  The HaluEval qa result requires a note. Its AUC is below 0.5 in output-level aggregation. It remains weak at the span
  level because the benchmark's correct answers are short named entities (typically three tokens: a single proper noun
  or quantity). In contrast, its hallucinated answers are fluent 15-token elaborations. Per-token mean log-delta is
  dominated by answer length and vocabulary register under this framing, not by hallucination-ness, and the primitive,
  as currently formulated, has nothing to contribute. This is a benchmark-structure artifact rather than a signal
  failure: HaluEval summarization and dialogue, which score generated text against fixed-length references, show the
  expected lifts.

  \subsubsection{True Entity-Level Classification}
  \label{sec:primitive-entity-level}

  Table~\ref{tab:span-aggregations} reports span-level classification results. A different and more granular question is
   whether the scoring primitive carries a usable signal at the level of individual extracted entities. This matters
  because many practical failure modes in summarization, retrieval-augmented generation, and agent outputs are
  concentrated in a small number of high-information spans: names of people, organizations, places, dates, and
  quantities.

  To test that setting directly, we ran an entity-level experiment on 500 FRANK summaries scored with Llama 3.1 8B. For
  each summary, spaCy NER was used to extract all named and numeric entities, yielding 985 entities across the sample.

  Each entity was then labeled in two ways. Under the strict label, an entity was treated as unsupported if its
  normalized surface form did not appear as a substring of the normalized source article. Under the relaxed label, an
  entity was treated as unsupported only if none of its content words appeared in the article. The relaxed label is
  narrower, but it more cleanly isolates cases where the generated output introduces genuinely new lexical material
  rather than a shortened form, paraphrase, or surface variation of something already present in the source.

  For each entity, we computed the mean over its BPE tokens of the same quantities used in
  Table~\ref{tab:span-aggregations}: $\log(P_{\text{LLM}}/P_{\text{RI}})$, rank-only
  $\log_2(r_{\text{global}}/r_{\text{source}})$, and the unconditional $-\log P_{\text{RI}}^{\text{global}}$, which is
  equivalent up to a constant to $\log_2(r_{\text{global}})$ under the Mandelbrot reference. Evaluation is reported as
  ROC-AUC per entity rather than per span.

  This setup lets us ask a narrower question than span-level detection: not whether a summary contains an error, but
  whether a particular extracted entity appears to be an unsupported high-information insertion.

  The results reveal a clear pattern. On the full entity set under the strict label (which includes paraphrases and
  abbreviations alongside genuine fabrications), all three features produce AUCs in the 0.37 to 0.38 range under the
  chosen score orientation, indicating no useful positive discrimination and suggesting either an inverted relation or a
   label-orientation mismatch for this slice. But when we restrict to named entities (PERSON, ORG, GPE) under the
  relaxed label, isolating cases where the model introduces genuinely new lexical material absent from the source,
  unconditional global rarity reaches AUC 0.622, and the other features reach 0.580 to 0.606. The strongest signal comes
   not from the most elaborate score but from the simplest: the unconditional global rarity term. Unsupported entities
  tend to be informationally heavy; they introduce rarer, more identity-bearing lexical items into a summary. The
  primitive is functioning less as a semantic fact-checker than as a lightweight specificity sensor for unsupported
  names, organizations, places, dates, and similar high-information spans.

\begin{table}[htbp]
  \centering
  \caption{True entity-level AUC on 500 FRANK summaries (Llama 3.1 8B, $\beta = 1$). Under the strict label on the full
  entity set, AUCs fall in the 0.37 to 0.38 range and indicate no useful positive discrimination; under the relaxed
  label on the named-entity subset (PERSON + ORG + GPE), unconditional global rarity reaches 0.622. The contrast
  reflects the primitive's role as a specificity sensor for unsupported high-information surface forms, not as a general
   entity fact-checker.}
  \label{tab:entity-auc}
  \begin{tabular}{lcc}
  \toprule
  \textbf{Feature} & \textbf{Strict, full set} & \textbf{Relaxed, PERSON+ORG+GPE} \\
                   & ($n=985$, 414 pos.)       & ($n=550$, 66 pos.)               \\
  \midrule
  $\log(P_{\text{LLM}}/P_{\text{RI}})$           & 0.373 & 0.606          \\
  $\log_{2}(r_{\text{global}}/r_{\text{source}})$ & 0.375 & 0.580          \\
  $-\log P_{\text{RI}}^{\text{global}}$          & 0.374 & \textbf{0.622} \\
  \bottomrule
  \end{tabular}
  \end{table}
  
  This entity-level signal is best understood as tracking whether the system has introduced lexically unsupported
  details, not whether it has correctly resolved the underlying entity relation. A method built on rank-frequency
  statistics should not be expected to recover referential correctness when the substitution involves entities of
  similar rarity; the taxonomy in Section~\ref{sec:taxonomy} already predicts this.

  That narrower interpretation is still practically valuable. In a compound evaluation pipeline, the entity-level
  primitive can serve as a cheap first-pass filter for unsupported high-information insertions, identifying which spans
  deserve escalation to more expensive downstream checks such as NLI, retrieval, or entity linking. In summarization
  systems, RAG pipelines, and agent frameworks, many operational failures arise not because every token is wrong, but
  because a few unsupported names, locations, or quantities are inserted with unwarranted confidence. The primitive is a
   detector of unsupported high-information surface forms, not a full verifier of entity truth.
 \subsubsection{Per-Domain $\beta$ Calibration: Estimating How Much to Trust the Baseline}
  \label{sec:beta-calibration}

  Section~\ref{sec:variational} introduced $\beta$ as a precision term recoverable from the variance of the
  rank-deviation signal on a domain-matched corpus:

  \begin{equation}
  \beta = \frac{1}{\sigma_{\Delta_{r}}^{2}}, \qquad \Delta_{r}(t_{i}) =
  \log_{2}\frac{r_{\text{global}}(t_{i})}{r_{\text{local}}(t_{i})}.
  \label{eq:beta-precision}
  \end{equation}

  The role of $\beta$ is best understood as a domain-specific trust coefficient on the shared reference distribution. If
   a domain lies close to the global Wikipedia baseline, deviations from that baseline are more informative and should
  be weighted more heavily. If a domain lies farther away, the same deviations should be discounted, because they are
  more likely to reflect ordinary register differences rather than genuine anomalies. On this reading, $\beta$ is not a
  hyperparameter to be swept for benchmark gain; it is a measured estimate of how much authority the shared baseline
  deserves in each domain.

  We estimate $\hat{\beta}_d$ in four steps. First, for each domain $d$, we assemble a held-out corpus disjoint from the
   scoring benchmarks: news from CC-News 2024, biomedical text from PubMed abstracts, legal text from CaseLaw Access
  Project excerpts, code from The Stack v2 Python, and social media from the PushShift Reddit dump. Second, we tokenize
  each corpus with the Llama 3.1 8B BPE vocabulary and compute the rank-deviation score $\Delta_{r}(t_{i})$ for every
  token occurrence, using the global token rank from the Wikipedia reference table and the local token rank from the
  domain corpus. Third, we compute the occurrence-weighted variance $\hat{\sigma}^{2}(\Delta_{r})$. Fourth, we set
  $\hat{\beta}_{d} = 1 / \hat{\sigma}^{2}(\Delta_{r})$. The resulting values are summarized in
  Table~\ref{tab:beta-domain}.

  \begin{table}[htbp]
  \centering
  \caption{Domain-level precision estimates for $\beta$ calibration. Higher $\hat{\beta}$ means that the domain more
  closely matches the shared reference distribution, so deviations from the baseline can be interpreted with greater
  confidence.}
  \label{tab:beta-domain}
  \begin{tabular}{lrrrr}
  \toprule
  \textbf{Domain} & \textbf{$N_{\text{occ}}$} & \textbf{mean $\Delta_{r}$} & \textbf{$\hat{\sigma}^{2}$} &
  \textbf{$\hat{\beta}$} \\
  \midrule
  News         & 13{,}025 & 2.05 &  8.62 & \textbf{0.116} \\
  Legal        & 14{,}327 & 2.37 &  9.78 & \textbf{0.102} \\
  Social media & 13{,}851 & 2.63 &  9.87 & \textbf{0.101} \\
  Biomedical   & 14{,}981 & 2.89 & 11.05 & \textbf{0.091} \\
  Code         & 20{,}694 & 5.10 & 19.52 & \textbf{0.051} \\
  \bottomrule
  \end{tabular}
  \end{table}

  Table~\ref{tab:beta-domain} shows that $\hat{\beta}$ varies by a factor of roughly 2.3 across domains. That spread is
  substantial enough to matter operationally. News, whose lexical distribution is closest to the Wikipedia reference,
  yields the highest precision. Code, whose subword structure and symbol mix differ most from running encyclopedic
  English, yields the lowest. The important point is not simply that domains differ; it is that the shared baseline is
  not equally trustworthy everywhere. $\hat{\beta}$ quantifies that difference in a single reusable coefficient.

  This interpretation becomes more useful once we notice that the scoring primitive caches the two sufficient statistics
   needed to recompute the posterior aggregate at any $\beta$: $\langle \log \Delta_{r} \rangle$ and $\langle -\log
  P_{\text{RI}} \rangle$. The aggregate can therefore be updated in closed form,

  \begin{equation}
  \langle \text{posterior}(\beta) \rangle = -\langle \log \Delta_{r} \rangle + (1 + \beta)\,\langle -\log P_{\text{RI}}
  \rangle,
  \label{eq:posterior-beta}
  \end{equation}

  without re-calling the LLM and without re-tokenizing the output. In deployment terms, that means calibration can be
  changed cheaply after the fact. Once the per-example statistics are cached, moving from $\beta = 1$ to a
  domain-matched $\hat{\beta}$ is a lightweight reweighting step rather than a second inference pass.

  Table~\ref{tab:beta-benchmark} reports what happens when each benchmark is rescored using its matched $\hat{\beta}$
  instead of the default $\beta = 1$.

  \begin{table}[htbp]
  \centering
  \caption{Benchmark-level AUC change under matched $\hat{\beta}$ relative to $\beta = 1$. The effect is modest but
  informative: domain-aware calibration does not uniformly improve scores, but it tends to help where the scored
  register is farther from the reference corpus.}
  \label{tab:beta-benchmark}
  \begin{tabular}{lccc}
  \toprule
  \textbf{Benchmark (task)} & \textbf{AUC at $\beta = 1$} & \textbf{AUC at matched $\hat{\beta}$} & \textbf{$\Delta$} \\
  \midrule
  FRANK (news)              & 0.557 & 0.529 & $-0.028$         \\
  TruthfulQA (social media) & 0.529 & 0.558 & \textbf{$+0.029$} \\
  HaluEval dialogue         & 0.520 & 0.535 & \textbf{$+0.015$} \\
  HaluEval summarization    & 0.519 & 0.524 & \textbf{$+0.005$} \\
  HaluEval qa               & 0.570 & 0.548 & $-0.022$         \\
  \bottomrule
  \end{tabular}
  \end{table}

  Matched $\beta$ helps on three of five slices and underperforms on two. That mixed pattern is not a weakness of
  interpretation; it is exactly what the calibration view predicts. If $\beta$ were merely a tunable scaling constant,
  one would expect a search for universal uplift. But if $\beta$ is a measured estimate of domain closeness to the
  reference, then its job is not to improve every benchmark. Its job is to prevent the baseline from being trusted
  equally in domains where it should not be.

  This is why the gains appear where they do. TruthfulQA, dialogue, and informal summarization are further from
  encyclopedic references and benefit from a more cautious weighting of baseline deviations. FRANK, by contrast, is
  news-like text that already lies close to the reference distribution; there, the equal-weighted default already
  performs competitively. HaluEval QA is dominated by benchmark structure and length effects, so no calibration of
  $\beta$ should be expected to rescue it. In other words, Table~\ref{tab:beta-benchmark} does not show a universal
  performance booster. It shows something more useful: a mechanism for making the scorer less brittle across
  heterogeneous registers.

  There is one clear caveat. The present analysis calibrates $\beta$ along a single axis: domain. Task type plausibly
  matters as well. QA, summarization, dialogue, and free-form continuation differ in valid response length, entity
  density, and lexical concentration, all of which may alter the effective precision of the rank-deviation signal. The
  values in Table~\ref{tab:beta-domain} should therefore be read as first-order domain calibrations, not as the final
  word on precision estimation.

  The operational conclusion is straightforward. $\beta$ is worth keeping not because it can be swept for leaderboard
  gains, but because it gives the scoring layer a principled way to adapt to domain shift without sacrificing its
  single-pass, CPU-only character. A lightweight domain detector, along with a cached $\hat{\beta}$ per domain, is
  sufficient to make the shared baseline more context-aware while preserving the primitive's low-cost design.

   \subsection{A Falsifiable Three-Tier Error Taxonomy}
  \label{sec:taxonomy}

  The application pattern above suggests a taxonomy that partitions the hallucination space by whether the error leaves
  a distributional signature.

  \begin{itemize}
      \item Tier 1, distributional anomaly. Unsupported surface-form entities, out-of-article entities, rare lexical
  insertions, and quantitative fabrications. These cases change which tokens appear in a way that the reference
  distribution can detect. Expected AUC range 0.60 to 0.95, depending on domain purity.
      \item Tier 1.5, mixed signal. Relational drift, circumstantial drift, and some vocabulary-preserving
  substitutions. Part of the error may surface as lexical or entity-level irregularity, but part remains in structure or
   referential interpretation. Expected AUC range 0.55 to 0.65.
      \item \textbf{Tier 2, world knowledge or reasoning.} Wrong claims expressed in unremarkable vocabulary. Scientific
   facts stated with appropriate scientific terminology, political falsehoods framed in conventional political language,
   coreference mistakes, and temporal or identity confusions whose tokens are all domain-appropriate. A distributional
  signal cannot see them. Expected AUC at chance.
  \end{itemize}

  The taxonomy is falsifiable. A finding that Tier 2 errors were reliably detected by any purely distributional method
  would refute it. Across the three benchmarks above, the taxonomy's predicted ordering holds: Tier 1 at AUC 0.60+, Tier
   1.5 at 0.55 to 0.60, Tier 2 at 0.50 and indistinguishable from chance ($p = 0.16$ on FRANK Tier 2).

  \subsection{Latency Profile}
  \label{sec:primitive-latency}

  Measured on a modest AMD Ryzen CPU, without GPU acceleration, on batches of 213 passages across three length bins with
   50 repetitions each, the primitive's scoring pass runs at:

  \begin{itemize}
      \item 0.139\,ms per passage with entities supplied, i.e.\ gap-only mode, which computes entity-gap scores against
  a pre-extracted list of entity spans and skips the NER pipeline on the hot path.
      \item 10.5\,ms per passage, including spaCy-based named entity recognition.
      \item approximately 2.6 microseconds per token, scaling linearly in output length.
  \end{itemize}

  One-time setup costs (rank table load and Mandelbrot normalization) total roughly 800\,ms and are amortized across all
   subsequent scoring.

  \textbf{Baselines.} We compare against two families of existing methods. Semantic Entropy [1], [2] generates $k$
  additional completions per example and estimates an entropy-like spread over semantic clusters; each extra completion
  requires a full LLM forward pass. SelfCheckGPT [3] is structurally similar but measures inter-sample consistency via
  an NLI model. Both reach AUC around 0.75 on the benchmarks relevant to us, but multiply the inference cost by $k$,
  which ranges from 5 to 20 in published configurations. White-box probes (Lookback Lens, Semantic Entropy Probes) are
  cheap but require access to attention maps or hidden states, which proprietary API models do not expose. The primitive
   in Section~\ref{sec:primitive-design} differs from all four on one axis: it requires no additional forward passes and
   no internals, at the cost of lower AUC on the same benchmarks.

\begin{figure}[htbp]
      \centering
      \includegraphics[width=0.95\textwidth]{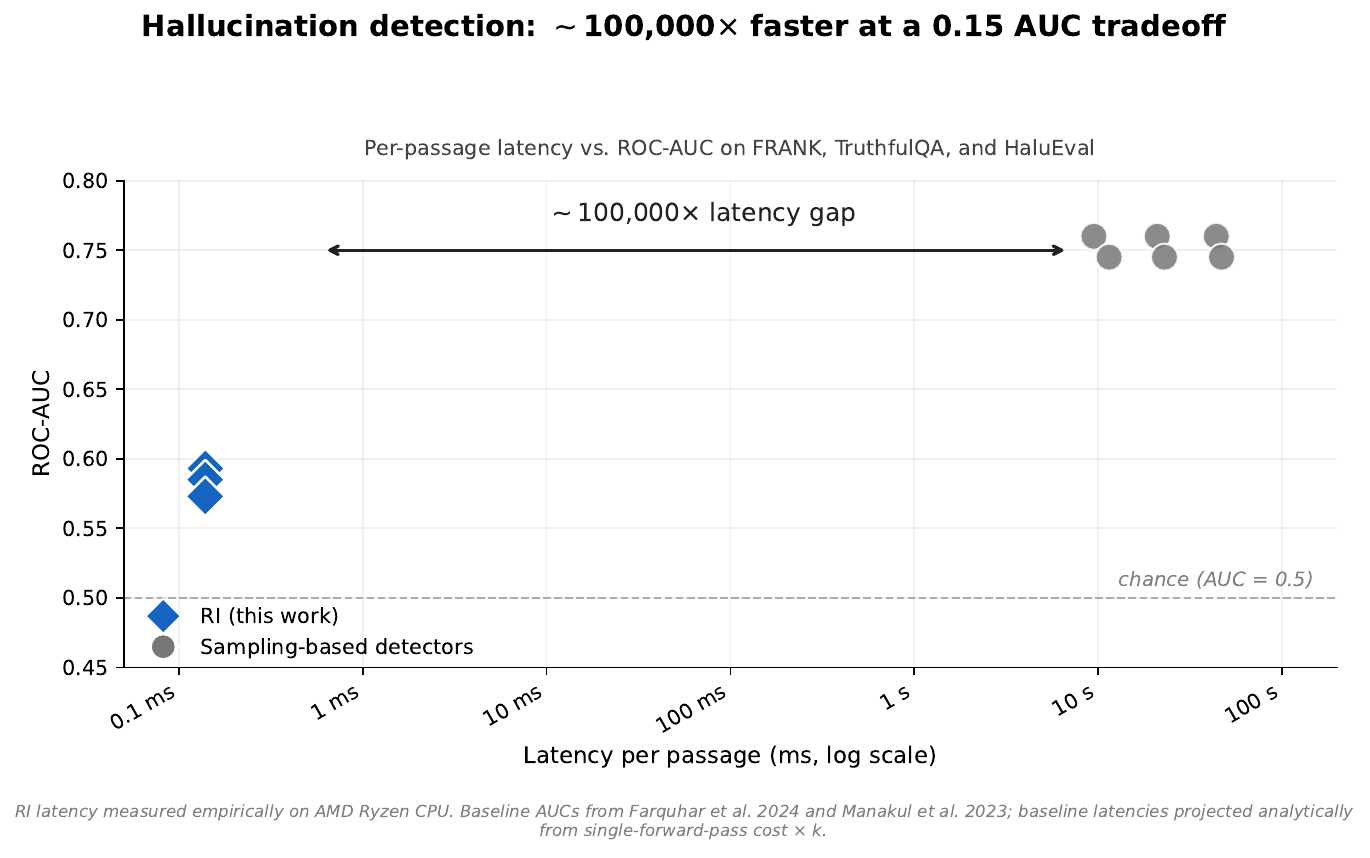}
      \caption{Latency-by-accuracy Pareto frontier. The primitive occupies the sub-millisecond region of the frontier at
   AUC $\approx 0.58$, while sampling-based detectors typically occupy the multi-second region at AUC $\approx 0.75$.
  The intended reading is stack position, not dominance: the primitive is designed as a first-pass triage layer.}
      \label{fig:latency-pareto}
  \end{figure}

  For the concrete comparison, sampling-based detectors at $k = 5$ typically require 20{,}000\,ms or more per passage on
   equivalent hardware. The primitive is therefore roughly 100,000$\times$ (five orders of magnitude) cheaper, at the AUC gap described above. This is a Pareto trade-off, not strict dominance in either direction (Figure~\ref{fig:latency-pareto}). The 2.6 microseconds per token figure is measured directly on the hardware described above. The sampling-based detector
  latencies (approximately 20{,}000\,ms per passage at $k = 5$) are order-of-magnitude estimates derived from published
  configurations rather than hardware-parity measurements on the same machine; a controlled same-hardware comparison is
  flagged as follow-up work in Section~\ref{sec:roadmap-followup}. The five-orders-of-magnitude gap is large enough that
   the Pareto-position conclusion is robust to reasonable variation in the baseline figure.

  \begin{figure}[htbp]
      \centering
      \includegraphics[width=0.95\textwidth]{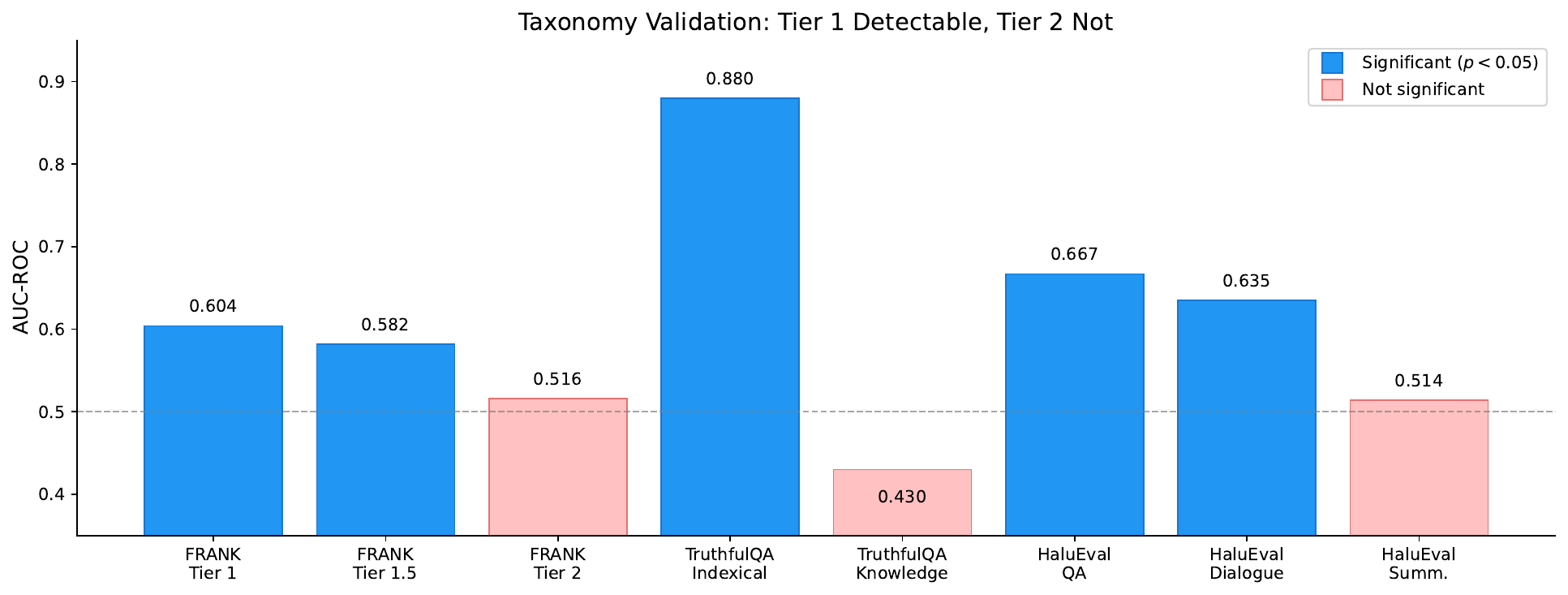}
      \caption{Taxonomy validation across benchmarks. The figure summarizes where the scoring layer is useful inside a
  larger system: Tier 1 anomalies are detectable above chance, mixed cases are weaker, and Tier 2 reasoning errors
  remain the job of complementary semantic verifiers.}
      \label{fig:taxonomy-validation}
  \end{figure}

  \subsection{Deployment Patterns}
  \label{sec:primitive-deployment}

  The primitive is most useful as the front end of a CPU-first assessment stack. A practical deployment pattern is:
  detect domain or task type; tokenize the output; run lightweight NER and number extraction; score tokens or spans
  against the reference rank table; optionally fuse in logprobs when they are available; and escalate only the uncertain
   or high-risk cases to heavier tools such as NLI, retrieval-based verification, semantic-entropy sampling, or
  tool-specific validators. Within that architecture, the following deployment patterns are especially attractive:

  \begin{itemize}
      \item \textbf{Cascade triage.} Route inputs to a more expensive detector based on the primitive's score. On FRANK,
   we observed that a configuration routing 10\% of inputs to an expensive detector, while scoring the remaining 90\%
  with the primitive alone, preserved most of the expensive-only baseline's accuracy at a 90\% reduction in expensive
  calls; the trade-off worsened rapidly at higher routing fractions. We report this as a preliminary observation on one
  benchmark. Production-scale deployability is not established by this number, and the primitive's structural weakness
  on reasoning errors (Section~\ref{sec:primitive-limits}) is a meaningful failure mode for any cascade routing on its
  scores.
      \item \textbf{Streaming token-by-token scoring.} At roughly 2.6 microseconds per token on CPU, the primitive fits
  inside a single token's generation budget. This enables real-time UI indication of suspect tokens and mid-generation
  interventions such as constrained decoding or resampling, neither of which is feasible with multi-sample detectors.
      \item \textbf{Tool call argument grounding.} Agent frameworks emit structured tool calls with typed arguments.
  Fabricated entities in those arguments are a canonical Tier 1 failure mode. A single primitive check on each tool
  call's arguments is cheap enough to run on every call and is targeted at the error type the primitive detects most
  reliably.
  \end{itemize}

  \subsection{What the Primitive Cannot Do}
  \label{sec:primitive-limits}

  The primitive is a distributional signal and shares all the limitations of distributional signals.

  \begin{itemize}
      \item It cannot see reasoning errors. A logically invalid derivation whose tokens are all plausible receives a
  normal score.
      \item It cannot see world knowledge errors in domain-appropriate vocabulary. A wrong statement of a well-known
  fact receives a normal score, because the primitive is contextually myopic: it reads each token's distributional
  properties against the reference corpus but does not read the surrounding context, the claim's factual content, or the
   relationships between entities in the passage.
      \item It is vulnerable to length artifacts on benchmarks where length and label are confounded.
      \item Its conviction does not guarantee accuracy reliably. Restricting attention to high-conviction scores
  increases AUC marginally but does not yield a high-accuracy subset.
  \end{itemize}

  These limitations are exactly why the primitive should be deployed as a component in compound systems. The shared
  reference distribution gives a useful first-pass signal on one axis of behavior. It does not resolve the broader
  assessment problem, nor does it justify claims of model interchangeability. We emphasize this point because the
  Mandelbrot framing, with its information-theoretic pedigree, can create an impression of deeper explanatory reach than
   the method actually delivers. The primitive is a distributional sensor. It detects surface-level statistical
  anomalies. It does not understand meaning, verify facts, or evaluate reasoning. Any deployment that treats it as a
  standalone verifier rather than as a triage layer will produce misleading confidence.

   \section{Limitations}
  \label{sec:limitations}

  We enumerate the limitations of both the convergence finding and the scoring primitive.

  \paragraph{On the convergence finding.}

  \begin{itemize}
      \item The six models measured so far (GPT-5.1, Claude Sonnet 4.6, Llama 3.1 8B, Gemini 2.5 Pro, Mistral Large,
  Qwen 2.5 7B) are representative of the current frontier and the 7-8B size class but not exhaustive.
  Sub-billion-parameter open-weight models and pre-trained-only base models remain to be characterized. 4-bit
  quantization is partially tested via the Q4\_K\_M Llama 3.1 8B disclosure in Section~3.1; more aggressive compression
  schemes (2-bit, pruning, distillation) remain open. All reported fits assume the shared-BPE normalization introduced
  in Section~3.4; whether fitted parameters and the magnitude of the fingerprinting separation are preserved under each
  vendor's native tokenizer is a tokenizer-invariance question that the present data are not designed to answer and that
   we scope as roadmap work (Section~8.1). We also do not report a head-to-head comparison of the two-parameter
  Mandelbrot form against a simpler unigram-frequency baseline scored directly from the reference corpus; the Mandelbrot
   form is motivated here by the variational derivation in Section~\ref{sec:variational} and by model selection against
  Zipf (Section~3.2), but whether the richer parameterization materially improves downstream scoring over a plain
  unigram reference is a question we flag as the first benchmark comparison to run in subsequent work.
      \item Coverage of long tail languages is limited. The convergence has been measured on English and code;
  multilingual Mandelbrot fits are a natural next step.
      \item Fit quality is lower in the head of the distribution (top 10 ranks). The coding theoretic cost model
  predicts this, but the practical consequence is that the most frequent tokens carry less diagnostic weight than the
  body of the distribution. Separately, the convergence finding is measured under vendor-default decoding policies
  (temperature 0.7, with top-p, top-k, seeds, and repetition penalties left at provider defaults rather than fixed
  across models; see Section~3.1). Whether the observed cross-vendor parameter stability persists under
  decoding-protocol parity is a methodological refinement we flag for follow-up work; the cross-vendor spread reported
  in Section~\ref{sec:fingerprinting} is nevertheless an order of magnitude larger than within-model bootstrap noise under
  the protocol actually used, so decoding variation would have to be a dominant driver of parameter differences to
  overturn the finding.
  \end{itemize}

  \paragraph{On the scoring primitive.}

  \begin{itemize}
      \item The primitive's raw AUC is below the best sampling-based detectors on the benchmarks tested. Its value lies
  in speed, portability, black-box applicability, and composability, not raw, standalone accuracy.
      \item The primitive's signal can be dominated by trivial confounds such as length on specific benchmarks.
      \item The primary pilot results use $\beta = 1$ as a simple baseline, while Section~\ref{sec:beta-calibration}
  reports an exploratory domain-matched $\beta$ measurement; the broader calibration story remains incomplete,
  especially along joint domain-by-task axes. The three pilot benchmarks (FRANK, TruthfulQA, HaluEval) are established
  but aging evaluation targets: TruthfulQA is partially saturated and may appear in some models' training data,
  HaluEval's QA split is dominated by a length artifact, and FRANK is summarization-specific. We chose this set because
  the three offer complementary error taxonomies against which the Section~\ref{sec:taxonomy} three-tier taxonomy makes
  specific, falsifiable predictions, but validation on more recent out-of-distribution benchmarks such as SimpleQA,
  RAGTruth, or HaluLens would materially strengthen the generalization claim and is a priority for follow-up work. The
  paper also does not include a head-to-head comparison with recent single-pass methods, such as Entropy Production Rate
   [31] or Layer-wise Semantic Dynamics [32], on identical benchmark splits; such a comparison would sharpen the Pareto
  characterization in Section~\ref{sec:primitive-latency}.
  \end{itemize}

   \section{Related Work and Positioning Within the Evaluation Stack}
  \label{sec:related}

  This paper sits at the intersection of statistical language regularities, hallucination detection, provenance
  analysis, and evaluation-systems design. The clearest way to position it is by stack layer: some methods characterize
  language itself, some trade latency for stronger black-box factuality signals, some exploit model internals, and some
  verify claims semantically against sources or knowledge bases. Our contribution targets a missing systems layer
  between generation and heavyweight verification: a cheap, analytic, model-agnostic first-pass scorer that can run on
  CPU and help decide which outputs deserve more expensive scrutiny.

  \paragraph{Rank-frequency laws in language.} Zipf [7] observed the empirical law; Mandelbrot [6] derived its
  information-theoretic form. Piantadosi [17] surveyed the broader debate on why such laws hold across languages. Cugini
   et al.\ [26] prove that local Zipf-Mandelbrot convergence is a universal property of ranked i.i.d.\ samples, a result
   that motivates our emphasis on parameter-level deviations rather than fit quality per se. Wang et al.\ [27]
  independently confirm that LLM-generated text obeys multiple linguistic scaling laws, including Zipf and Heaps, with
  exponents close to human text, and use the observation for data augmentation rather than output scoring. Their use of
  the label ``Mandelbrot's law'' refers to multifractal detrended fluctuation analysis, not to the classical
  rank-frequency distribution fitted in this paper; the two share a name but are distinct objects. Manin [28] notes that
   Mandelbrot's original derivation constrains $q$ and $s$ to co-vary, limiting the effective dimensionality of the
  fingerprint space. Our contribution is not to re-establish that LLM outputs follow these laws (this is now
  independently documented) but to treat the regularity as deployable infrastructure for per-output, CPU-only assessment
   and to derive a concrete scoring primitive from it.

  \paragraph{Sampling-based hallucination detectors.} Semantic Entropy [1], [2] and SelfCheckGPT [3] probe model
  self-consistency by repeated sampling. They can deliver materially stronger per-example factuality signals than our
  primitive ones, but they do so by paying for multiple extra completions or auxiliary inference passes. In stack terms,
   they are excellent second-stage verifiers when the latency budget allows; they are not natural first-pass filters for
   very high-volume CPU-bound deployments.

  \paragraph{White-box probes.} Lookback Lens [4] reads attention maps; Semantic Entropy Probes [5] train linear
  classifiers on hidden states. These approaches are efficient inside controlled deployments, but they require access to
   internals that proprietary API models do not expose. Their natural home is model-side monitoring, not vendor-agnostic
   assessment across heterogeneous closed and open systems.

  \paragraph{Combining neural predictions with distributional statistics and contrastive decoding.} Nikkarinen et al.\
  [18] combine neural model predictions with unigram frequencies using a product-of-experts architecture for language
  modeling. Our posterior is formally similar but is used at inference time on arbitrary model outputs, uses rank
  deviation rather than raw unigram counts, and derives $\beta$ as a precision term rather than setting it
  heuristically. Li et al.\ [19] use contrastive decoding to compare strong and weak models; our primitive instead
  substitutes an analytic reference distribution, eliminating the need for a second model or a second forward pass.

  \paragraph{Lexical overlap, NLI, and source-conditioned verification.} ROUGE [20] remains useful on summarization
  faithfulness when lexical overlap is informative. NLI-based approaches [21] are better suited to reasoning errors,
  entailment failures, and vocabulary-preserving contradictions that our primitive cannot see. In systems terms, these
  are natural complementary stages: the distributional primitive is a cheap gate, whereas source-conditioned semantic
  verification is the heavier layer that should follow on the hard cases.

  \paragraph{Hallucination taxonomies, provenance, and watermarking.} Surveys by Ji et al.\ [22], Huang et al.\ [23],
  and Zhang et al.\ [24] map the hallucination landscape. Our three-tier taxonomy differs in being operational rather
  than editorial: it partitions errors by whether they should leave a detectable distributional trace. Concerning
  provenance, Kirchenbauer et al.\ [25] propose cryptographic watermarking, whereas our fingerprinting proposal is
  statistical and model-agnostic. McGovern et al.\ [29] demonstrate that categorical frequency histograms over POS tags,
   constituency labels, and top-k tokens serve as passive model-family fingerprints; their method uses a fixed small
  tagset and a trained classifier, whereas ours parameterizes the full rank-frequency tail with two continuous scalars
  and requires no training. Both approaches belong to the same governance space but at different levels of abstraction
  and guarantee.

  Taken together, prior work leaves an operational gap. Sampling-based methods are too expensive for universal
  deployment. White-box probes do not apply to closed APIs. Recent single-pass methods (EPR, LOS-Net) require either API
   logprobs or learned parameters. To our knowledge, no published method offers a fully analytic, zero-logprob, CPU-only
   scoring signal that is cheap enough to run on every output, honest enough to acknowledge its limits, and still useful
   enough to guide a larger evaluation pipeline. That is the gap this paper aims to fill.

  \section{Systems Roadmap and Future Directions}
  \label{sec:roadmap}

  The next stage is not simply to accumulate more fit plots. It is to turn the shared rank-frequency signature into a
  dependable evaluation component that can live inside real agent and LLM stacks. We therefore organize the roadmap into
   four tracks: empirical consolidation of the shared signature, hardening of the scoring layer itself, stack
  integration, and validation of the secondary auditing uses the paper sketches.

  \subsection{Consolidate the empirical foundation}
  \label{sec:roadmap-foundation}

  The first track is to stress-test the shared signature itself: broaden the cohort to smaller models, base models, more
   aggressive compression schemes, multilingual settings, and different reference-corpus refresh intervals. The goal is
  not rhetorical universality but boundary mapping: identify where the Mandelbrot baseline remains stable enough to be
  operationally reused, and where it breaks or must be specialized.

  \subsection{Harden calibration and benchmark alignment}
  \label{sec:roadmap-followup}

  The second track is to harden the scoring primitive as a component. That means improving $\beta$ calibration along
  joint domain-by-task axes; releasing reusable reference-rank artefacts and lightweight tokenization/NER tooling; and
  running matched head-to-head comparisons against sampling-based and source-conditioned baselines on the same benchmark
   slices. This is where the method either earns or loses its place in a serious evaluation stack.

  \subsection{Build the primitive into evaluation stacks}
  \label{sec:roadmap-stack}

  The third track is systems integration. The most natural experiments are cascade routing, streaming monitoring,
  tool-call argument grounding, and training-health dashboards. Here, the question is no longer ``does the statistic
  exist?'' but ``how much downstream cost, latency, and risk can it remove when used as a first-pass filter before
  heavier verifiers?''

  \subsection{Validate provenance, auditing, and governance uses}
  \label{sec:roadmap-governance}

  A fourth near-term workstream concerns the secondary applications that become plausible once the baseline is accepted
  as infrastructure: statistical model fingerprinting, synthetic-text auditing, contamination diagnostics, and
  cross-model consensus scoring. These are promising precisely because the reference is model-agnostic, but they require
   their own false-positive studies, adversarial robustness tests, and task-specific evaluation protocols before they
  can be operationalized.

  Across all four tracks, the paper's thesis stays the same: the value of the shared signature is not that it replaces
  semantic verification, but that it may furnish the missing first stage of a CPU-first assessment stack. The decisive
  evidence will therefore come from integrated system evaluations, not from fit quality alone.

   \section{Conclusion}
  \label{sec:conclusion}

  The empirical finding reported here has two parts, and both matter. The first is a surprising family-level
  convergence: on the token rank-frequency axis, outputs from the measured six-model frontier cohort are uniformly well
  described by the same two-parameter Mandelbrot form, with $R^{2} > 0.94$ in 34 of 36 conditions and cross-vendor
  parameter stability across six independently trained models. Section~3.3 positions this finding relative to the local
  universality theorem of Cugini et al.\ [26] and the independent scaling-law observations of Wang et al.\ [27]; the
  three empirical properties the theorem does not predict in our setting (conformance despite non-i.i.d.\ autoregressive
   generation, global rather than piecewise fit across more than three decades of rank, and cross-vendor parameter
  clustering) are the empirical content of the finding. The second part, and the more practically consequential, is that
   the shared family does not collapse the models into statistical duplicates. Fitted parameters remain cleanly
  separable, with cross-model spread exceeding within-model bootstrap noise by more than an order of magnitude. Nothing
  in the universality theorem predicts separation, and it is this parameter-level finding, together with the
  shared-family finding, that supports the two downstream capabilities: a reusable Mandelbrot baseline for black-box
  output assessment, and a statistical fingerprinting signal that can test whether a vendor-delivered output is
  consistent with its claimed model family.

  The value of this infrastructure lies in its architecture. The scoring primitive is cheap enough to run on every
  output, modest enough to remain honest about what it cannot detect, and useful enough to steer downstream computation
  toward cases that deserve more expensive verification. The fingerprinting signal is statistical rather than
  cryptographic, but it does not require provider cooperation, watermark embedding, or access to model internals; it
  complements those mechanisms in a governance stack rather than replacing them. Both applications sit upstream of
  semantic verification and downstream of raw generation, occupying a stack position that existing methods have not
  filled. The decisive next step is integrated system evaluation: measuring how much downstream cost, latency, and risk
  each of these signals can remove when used within real agent and LLM pipelines, rather than the further accumulation
  of fit quality across new domains.

  \section*{Competing Interests}

  The authors are principals of Evolutionairy AI, and of Castle Ridge Asset Management, which is a shareholder
  in Evolutionairy AI. A provisional patent application is in preparation covering the methods described in this paper.

  \section*{Acknowledgments}

  We thank Ibrahim Shaer, Anas Ibrahim, and Tamaz Andguladze of Evolutionairy AI for detailed manuscript review, framing
   recommendations, and contributions to the prior art survey that shaped the paper's positioning within the evaluation,
   fingerprinting, and scaling-laws literature. Their input materially improved the clarity and scope of the work.

  \section*{Data and Code Availability}

  A reference implementation of the scoring primitive, the precomputed Mandelbrot rank table built from the
  4-billion-token Wikipedia snapshot described in Section~2.2, fitting and evaluation scripts for the benchmarks
  reported in Section~\ref{sec:primitive}, and configurations sufficient to reproduce Tables~\ref{tab:fit-quality}, \ref{tab:fingerprint}, \ref{tab:span-aggregations}, \ref{tab:entity-auc},
  \ref{tab:beta-domain}, and~\ref{tab:beta-benchmark} are released
  at: \url{https://github.com/Evolutionairy-AI/Ranking-Inference} under an open-source license. Prompts and model outputs used for the
  fits in Section~3 are released alongside the code. Proprietary API outputs are released to the extent permitted by
  each provider's terms of service; where redistribution is restricted, we release the fitted rank tables and the
  scripts needed to regenerate outputs from the same prompts.

\end{document}